\documentclass[pra,aps,10pt,twocolumn,floatfix,superscriptaddress, notitlepage,longbibliography]{revtex4-2}

\usepackage[T1]{fontenc}
\usepackage[utf8]{inputenc}
\usepackage{xcolor}
\usepackage{graphicx}
\usepackage{amsmath,amssymb,amsfonts, amsthm}
\usepackage{quantikz}
\usepackage{bm}
\usepackage{braket}
\usepackage{tikz}
\usepackage{booktabs}
\usepackage{multirow}
\usepackage{siunitx}
\usepackage{eurosym}
\usepackage{array}

\usepackage[colorlinks=true,citecolor=blue,linkcolor=magenta,urlcolor=blue]{hyperref}
\expandafter\def\expandafter\UrlBreaks\expandafter{\UrlBreaks\do\-}

\newcommand{\E}{\mathbb{E}}
\newcommand{\mmd}{\operatorname{MMD}}
\newtheorem{proposition}{Proposition}

\begin{document}

\title{Classical training of conditional instantaneous quantum polynomial-time Born machines}

\author{Oriol Balló-Gimbernat}
\email{oriol.ballo@eurecat.org}
\affiliation{Eurecat, Centre Tecnològic de Catalunya, Barcelona, Spain}
\affiliation{Centre de Visió per Computador (CVC), Barcelona, Spain}
\affiliation{Universitat Autònoma de Barcelona (UAB), Barcelona, Spain}

\author{Emiliano Tolotti}
\affiliation{Department of Information Engineering and Computer Science, University of Trento, Via Sommarive 9, 38123 Trento, Italy}

\author{Laia Domingo}
\affiliation{Centre de Visió per Computador (CVC), Barcelona, Spain}

\author{Fernando Vilariño}
\affiliation{Centre de Visió per Computador (CVC), Barcelona, Spain}
\affiliation{Universitat Autònoma de Barcelona (UAB), Barcelona, Spain}

\author{Adan Garriga}
\affiliation{Eurecat, Centre Tecnològic de Catalunya, Barcelona, Spain}

\date{\today}

\begin{abstract}
Instantaneous quantum polynomial-time (IQP) Born machines are attractive quantum generative models, as certain IQP circuit families are conjectured to be hard to sample, yet they can be trained classically.  Previously proposed conditional IQP Born machines make the parameters of the diagonal layer condition dependent. We introduce conditional IQP Born machines, in which the condition is encoded in the input state and a single IQP unitary is shared across conditions. Efficient sampling and amplitude-ratio evaluation for the reference state entering the diagonal layer suffice to obtain unbiased classical estimates of a conditional maximum mean discrepancy objective and its gradients. We show that input conditioning can represent conditional families that no diagonally conditioned IQP model with the same number of qubits can reproduce. For the sixteen-qubit Curie--Weiss spin model, a single conditional model captures the magnetization transition at held-out temperatures. Its mean total variation distance is six times lower than that of diagonal conditioning under the same protocol and comparable to that of classical generative models with up to seven times as many parameters. We also train conditional models classically with up to $2048$ qubits, which reproduce the mean squared magnetization of the target away from the crossover. These results extend classically trained IQP Born machines to conditional generation.
\end{abstract}
\maketitle
\raggedbottom

\begin{figure*}[t]
\centering
\includegraphics[width=\textwidth]{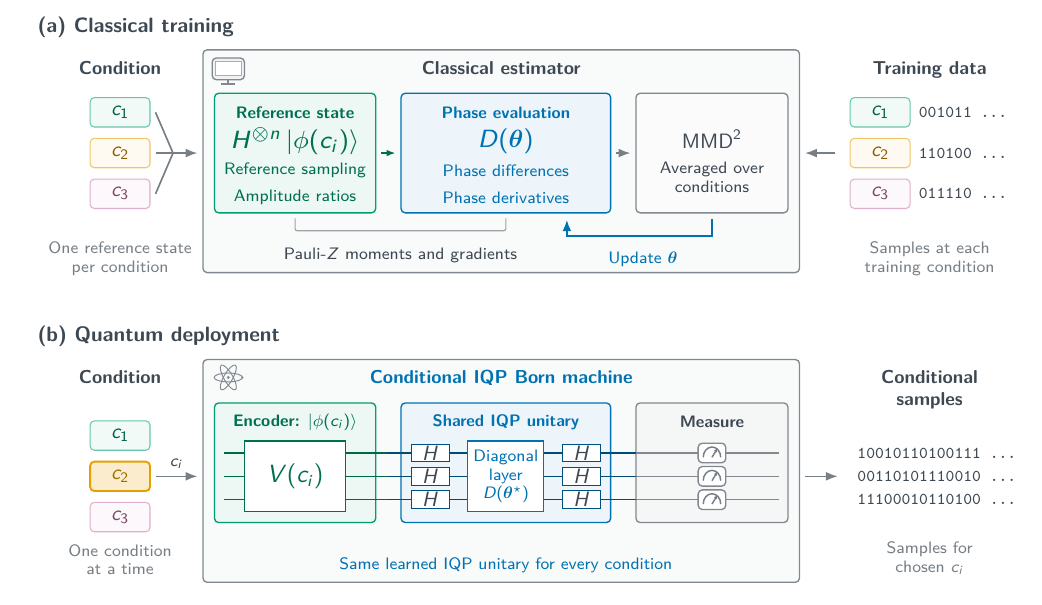}
\caption{Proposed framework for classical training and quantum inference of conditional instantaneous quantum polynomial-time (IQP) Born machines. An encoder $V(c_i)$ maps each condition $c_i$ to an input state $\ket{\phi(c_i)}$. (a) Classical training. Reference samples and amplitude ratios from $H^{\otimes n}\ket{\phi(c_i)}$ are combined with phase differences and derivatives from the diagonal layer $D(\boldsymbol\theta)$ to estimate conditional Pauli-$Z$ expectation values and their gradients. These estimates update the parameters $\boldsymbol\theta$ by minimizing the squared maximum mean discrepancy ($\mathrm{MMD}^2$) averaged over the training conditions. (b) Quantum deployment. After training, the parameters $\boldsymbol\theta^\star$ define the quantum circuit. Applying the trained IQP unitary, shared across all conditions, to $\ket{\phi(c_i)}$ and measuring in the computational basis yields samples from the learned conditional distribution $q_{\boldsymbol\theta^\star}(\cdot\mid c_i)$.}
\label{fig:pipeline_workflow}
\end{figure*}

\section{Introduction}
Conditional generative models approximate a family of target distributions $p(\cdot\mid c)$ with a single model. They capture how a distribution depends on one or more conditioning variables, such as the temperature of a physical system or a class label in image generation. Learning across conditions aims to capture structure shared by the family, which in turn supports generation at conditions not encountered during training. In classical machine learning, examples include conditional variational autoencoders \cite{sohnConditionalGenerative2015, walkerUncertainFutureForecasting2016, yanAttribute2ImageConditionalImage2016}, generative adversarial networks \cite{mirzaConditionalGAN2014, miyatoCGANsProjectionDiscriminator2018, isolaImagetoImageTranslationConditional2018}, and diffusion models \cite{hoClassifierFreeDiffusionGuidance2022, dhariwalDiffusionModelsBeat2021}. Among quantum generative models, a widely studied architecture is the quantum circuit Born machine (QCBM), which represents a probability distribution through the measurement statistics of a parameterized circuit \cite{benedettiGenerativeModeling2019,liuDifferentiableLearning2018}. QCBMs have been extended to conditional generation through input-state preparation, conditional registers, and postselection \cite{heConditionalGenerativeModel2019,kissConditionalBornMachine2022,zengConditionalQCBM2023,zengLearningInferenceGenerative2019,certoConditionalGenerativeModels2023}.

Training QCBMs at scale requires quantum hardware, as the cost of classical simulation generally grows exponentially. However, training on hardware is also costly. Gradients are typically estimated with the parameter-shift rule, which requires a number of circuit evaluations that grows linearly with the number of parameters \cite{wierichsGeneralParametershiftRules2022}. Circuits with commuting generators reduce this overhead to backpropagation-like scaling \cite{bowlesBackpropagationScaling2025}, but training on noisy intermediate-scale quantum hardware \cite{preskillQuantumComputingNISQ2018} still faces noise-induced optimization issues, such as barren plateaus \cite{wangNoiseinducedBarrenPlateaus2021}.

Instantaneous quantum polynomial-time (IQP) circuits \cite{shepherdInstantaneousQuantumComputation2009} offer a way around some of these limitations. They are attractive architectures for QCBMs \cite{coyleBornSupremacyQuantum2020}, because sampling from certain IQP circuit families is conjectured to be classically hard under standard complexity-theoretic assumptions \cite{bremnerClassicalSimulationCommuting2011,bremnerAverageCaseComplexityApproximate2016,bremnerAchievingQuantumSupremacy2017,marshallImprovedSeparationQuantum2024}, yet their Pauli-$Z$ expectation values can be estimated classically to additive accuracy \cite{nestSimulatingQuantumComputers2011}. Objectives expressed through these expectation values, such as the maximum mean discrepancy (MMD) with a Gaussian kernel \cite{grettonKernelTwoSample2012,rudolphTrainabilityBarriersOpportunities2024}, can therefore be estimated and optimized classically for circuits far beyond the reach of exact simulation \cite{rudolphTrainabilityBarriersOpportunities2024,recio-armengolTrainClassicalDeploy2026}. Recio-Armengol \textit{et al.} combined these ingredients into a framework in which IQP Born machines are trained classically and require quantum hardware only for sampling \cite{recio-armengolTrainClassicalDeploy2026}, an approach since extended to non-binary discrete data \cite{huangSpectralBorn2026}.

In this framework, the IQP unitary acts on a fixed input state, typically $\ket{0}^{\otimes n}$, and represents a single distribution. Existing conditional IQP Born machines make part of the diagonal layer condition dependent. Liu \textit{et al.} train a circuit core at a reference condition, freeze it, and then adapt a block of low-order diagonal parameters sequentially at each further condition, interpolating the adapted blocks to unseen conditions \cite{liuCorrelationComplexity2026,liuPhysicsInformedLatent2026}. ConQuER instead fits a separate controller layer of diagonal gates, added to a pretrained generator, to data filtered by each target condition \cite{zouConQuER2025}. In both approaches, no parameters are trained jointly across conditions, and the dependence on the condition is not learned from the family as a whole. Outside the IQP setting, Kiss \textit{et al.} conditioned a QCBM through its input state and trained a single circuit across conditions with an MMD loss and gradients estimated with the parameter-shift rule from circuit samples \cite{kissConditionalBornMachine2022}. Conditioning the input state while training a shared IQP unitary jointly across conditions and estimating the objective and its gradients classically has not been studied.

In this work, we encode the condition in the input state of an IQP Born machine and share a single IQP unitary across conditions, so that the dependence on the condition is learned from the whole family (Fig.~\ref{fig:pipeline_workflow}). This extends the framework of Recio-Armengol \textit{et al.} to condition-dependent inputs while retaining classical training \cite{recio-armengolTrainClassicalDeploy2026}. The underlying estimators require only sampling and amplitude ratios of the reference state, the input after the first Hadamard layer of the IQP unitary \cite{nestSimulatingQuantumComputers2011,havlicekAmplitudeRatiosNeural2023}, so the encoder may include non-diagonal operations while the trainable block remains an IQP unitary (Sec.~\ref{sec:model-access}). Matrix product and permutation-symmetric states, for example, support both operations efficiently.

Input conditioning also enlarges the conditional families that the model can represent. An IQP unitary cannot change the output's distribution in the $X$ basis, which is therefore fixed by the input state. As a result, input conditioning can represent families that diagonal conditioning on a fixed input cannot reproduce with the same number of qubits (Sec.~\ref{sec:gauge}, Proposition~\ref{prop:separation}), and it can impose output symmetries, such as global spin-flip invariance, on every member of the family (Sec.~\ref{sec:input-symmetry}).

We test the framework at sixteen qubits on two target families, with diagonal conditioning as the main baseline (Secs.~\ref{sec:targets}--\ref{sec:results}). On a family of Hamming-weight distributions with tunable mean and width, a model trained with each coordinate varied separately generalizes to their combinations, with about one third of the total variation distance (TVD) of diagonal conditioning. On the Curie--Weiss spin model, a single model captures the transition from a unimodal to a bimodal magnetization profile at held-out temperatures, and the encoder matters: an encoder that conditions only through input-state phases already lowers the TVD $2.7$ times relative to diagonal conditioning, and adding a one-axis-twisting step lowers it more than twofold further, to one sixth of diagonal conditioning. This shared model is, on average, as accurate as circuits trained at each evaluation temperature and comparable to classical generative models with up to seven times as many parameters.

Because Curie--Weiss observables remain classically tractable at any size, we also train conditional models classically with up to $2048$ qubits and check that they reproduce the mean squared magnetization away from the crossover. Finally, we sample models with up to $30$ qubits on real quantum hardware (Appendix~\ref{sec:hardware}).

\section{Conditional IQP Born machines}
\label{sec:cond_model}

Consider a target family of distributions $\{p(\cdot\mid c):c\in\mathcal C\}$ over $n$-bit strings, which we model using $n$ qubits. A conditional IQP Born machine consists of two components: an encoder that maps each condition $c$ to an input state $\ket{\phi(c)}$, and a parameterized IQP unitary $U_{\boldsymbol{\theta}}$ with parameters $\boldsymbol\theta$. The model represents the distribution family as
\begin{equation}
    q_{\boldsymbol{\theta}}(x\mid c)
    =\left|\bra{x}U_{\boldsymbol{\theta}}\ket{\phi(c)}\right|^2,
    \qquad x\in\{0,1\}^n .
    \label{eq:conditional-model}
\end{equation}
Training aims to find a single unitary $U_{\boldsymbol{\theta}}$ that maps each input state $\ket{\phi(c)}$ to a distribution $q_{\boldsymbol{\theta}}(\cdot\mid c)$ that closely approximates the target $p(\cdot\mid c)$ for all $c\in\mathcal{C}$.

The IQP unitary has the form
\begin{equation}
    U_{\boldsymbol{\theta}}
    =H^{\otimes n}D(\boldsymbol{\theta})H^{\otimes n},
    \label{eq:iqp-unitary}
\end{equation}
where the diagonal layer
\begin{equation}
    D(\boldsymbol\theta)=\exp\Bigl(i\sum_{g\in\mathcal G}\theta_g\,g\Bigr)
    \label{eq:graph-diagonal}
\end{equation}
is parameterized by a phase $\theta_g$ for each generator in a set $\mathcal G$ of polynomially many Pauli-$Z$ strings of length $n$. Because the Hadamard layers exchange the $Z$ and $X$ bases, $U_{\boldsymbol\theta}$ is diagonal in the $X$ basis. The diagonal layer acts on the reference state $\ket{\psi_c}=H^{\otimes n}\ket{\phi(c)}$, whose amplitudes we write $\psi_c(z)$. We include the initial Hadamard layer in $U_{\boldsymbol\theta}$ so that the input $\ket{0}^{\otimes n}$ recovers the standard IQP Born machine.

\subsection{Sampling-and-ratio interface}
\label{sec:model-access}

Classical training of a standard IQP Born machine relies on estimating the Pauli-$Z$ moments of its output distribution,
\begin{equation}
    \mu_q(a)
    =\bra{0}^{\otimes n}U_{\boldsymbol\theta}^{\dagger}Z_aU_{\boldsymbol\theta}\ket{0}^{\otimes n},
    \qquad
    Z_a=\bigotimes_{i=1}^{n}Z_i^{a_i},
    \label{eq:pauli-moment}
\end{equation}
for bit masks $a\in\{0,1\}^n$ \cite{nestSimulatingQuantumComputers2011,recio-armengolTrainClassicalDeploy2026}. Van den Nest's sparse-observable method estimates such moments for computationally tractable reference states \cite{nestSimulatingQuantumComputers2011}, and Havl\'{i}\v{c}ek showed that two operations suffice: sampling the computational-basis distribution and evaluating ratios of amplitudes \cite{havlicekAmplitudeRatiosNeural2023}. For the standard reference state $\ket{+}^{\otimes n}$, the samples are uniform bitstrings and all ratios equal one, which gives the cosine estimator of IQP moments \cite{nestSimulatingQuantumComputers2011,recio-armengolIQPoptFastOptimization2025,recio-armengolTrainClassicalDeploy2026}.

Because the estimator accesses the reference state only through these two operations, any condition-dependent reference state that supports them efficiently admits the same estimation; this is what lets us move the condition from the diagonal layer into the input state. Examples of reference states that support both operations efficiently include product states, matrix product states of polynomial bond dimension \cite{nestSimulatingQuantumComputers2011}, and permutation-symmetric states, whose amplitudes depend only on the Hamming weight. For each condition $c$, we require exact samples $z\sim|\psi_c(z)|^2$ and, for a sampled $z$ and a mask $a$, the amplitude ratio
\begin{equation}
    R_c(z,a)=\frac{\psi_c(z\oplus a)}{\psi_c(z)},
    \label{eq:amplitude-ratio}
\end{equation}
where $\oplus$ denotes bitwise addition modulo two. We call these two operations the sampling-and-ratio interface of the reference state. Replacing $\ket{0}^{\otimes n}$ by $\ket{\phi(c)}$ in Eq.~\eqref{eq:pauli-moment} defines the conditional moments $\mu_q(a\mid c)$. With the sampling-and-ratio interface, these moments, the MMD objective of Sec.~\ref{sec:cond_training} built from them, and their gradients can be estimated to additive accuracy $\epsilon$ in time polynomial in $n$ and $1/\epsilon$ (Appendix~\ref{sec:estimator-variance}).

\subsection{Classical estimation of Pauli-\texorpdfstring{$Z$}{Z} moments}
\label{sec:ratio-estimator}

The diagonal layer multiplies each computational-basis state by a phase, $D(\boldsymbol\theta)\ket{z}=e^{if_{\boldsymbol\theta}(z)}\ket{z}$, where
\begin{equation}
    f_{\boldsymbol\theta}(z)=\sum_{g\in\mathcal G}\theta_g\chi_g(z),
    \qquad
    \chi_a(z)=(-1)^{a\cdot z},
    \label{eq:iqp-phase-function}
\end{equation}
and each generator is identified with its support $g\in\{0,1\}^n$, so that $\chi_g(z)$ is its eigenvalue on $\ket z$. Each moment is an expectation over the reference distribution,
\begin{equation}
    \mu_q(a\mid c)
    =\E_{z\sim|\psi_c|^2}\!\left[R_c(z,a)\,e^{i\Delta_af_{\boldsymbol\theta}(z)}\right],
    \label{eq:moment-as-reference-average}
\end{equation}
where $\Delta_af_{\boldsymbol\theta}(z)=f_{\boldsymbol\theta}(z\oplus a)-f_{\boldsymbol\theta}(z)$ and $R_c$ is the ratio of Eq.~\eqref{eq:amplitude-ratio}. This holds because the Hadamard layers turn $Z_a$ into the bit flip $X_a$, which pairs each string $z$ with $z\oplus a$. Since strings with $\psi_c(z)=0$ are never sampled, the ratio is needed only where it is defined. Equation~\eqref{eq:moment-as-reference-average} is Havl\'{i}\v{c}ek's amplitude-ratio estimator \cite[Thm.~5]{havlicekAmplitudeRatiosNeural2023}, built on Van den Nest's sparse-observable method \cite[Thm.~3]{nestSimulatingQuantumComputers2011}, applied to the one-sparse operator $D(\boldsymbol\theta)^\dagger X_aD(\boldsymbol\theta)$, whose expectation in $\ket{\psi_c}$ is $\mu_q(a\mid c)$. The ratios do not depend on the normalization of $\psi_c$.

Since the moment is real and only the phase factor depends on $\boldsymbol\theta$,
\begin{equation}
\begin{aligned}
    Y_{c,a}(z;\boldsymbol\theta)&=\operatorname{Re}\bigl[R_c(z,a)\,e^{i\Delta_af_{\boldsymbol\theta}(z)}\bigr],\\
    Y'_{g,c,a}(z;\boldsymbol\theta)&=\operatorname{Re}\bigl[i\,d_{g,a}(z)\,R_c(z,a)\,e^{i\Delta_af_{\boldsymbol\theta}(z)}\bigr]
\end{aligned}
    \label{eq:ratio-sample}
\end{equation}
are unbiased single-sample estimators of $\mu_q(a\mid c)$ and $\partial_{\theta_g}\mu_q(a\mid c)$. Here, the derivative of $\Delta_af_{\boldsymbol\theta}$ with respect to $\theta_g$, $d_{g,a}(z)=\chi_g(z\oplus a)-\chi_g(z)$, equals $-2\chi_g(z)$ when the supports of $g$ and $a$ overlap in an odd number of positions and zero otherwise. Averaging $S$ samples gives mean-square errors of at most $1/S$ and $4/S$, without any bound on individual ratios \cite[Sec.~3.4]{havlicekAmplitudeRatiosNeural2023}, because $|d_{g,a}|\leq2$ and
\begin{equation}
    \E_{z\sim|\psi_c|^2}|R_c(z,a)|^2=\sum_{z:\psi_c(z)\neq0}|\psi_c(z\oplus a)|^2\leq1.
\end{equation}

\subsection{Encoders beyond diagonal conditioning}
\label{sec:gauge}

Because $U_{\boldsymbol\theta}$ is diagonal in the $X$ basis, it multiplies each reference amplitude $\psi_c(z)$ by a phase and leaves its magnitude unchanged. For every parameter vector, the output's $X$-basis distribution therefore equals the reference distribution,
\begin{equation}
    \bigl|\bra{z}H^{\otimes n}U_{\boldsymbol\theta}\ket{\phi(c)}\bigr|^2=|\psi_c(z)|^2.
    \label{eq:x-basis-conservation}
\end{equation}
The encoder thus influences the output in two ways: through the phases of the reference amplitudes, which a diagonal layer can also modify, and through their magnitudes, which no diagonal layer can alter. Diagonal conditioning on a fixed input changes the output only through the phases and keeps the same $X$-basis distribution at every condition.

Although the model samples in the computational basis, this conservation constrains its output, because the computational-basis and $X$-basis distributions of a pure state are not independent. For example, $\ket{0}^{\otimes n}$ has a deterministic computational-basis outcome and a uniform $X$-basis distribution, and $\ket{+}^{\otimes n}$ the reverse. We call an $X$-basis distribution \emph{compatible} with an output distribution $p$ if some pure state has both, and say that a family has a \emph{common compatible} $X$-basis distribution if one distribution is compatible with all its members.

\begin{proposition}
\label{prop:separation}
Consider IQP Born machines on $n$ qubits without ancillas. (i) Diagonal conditioning on a fixed pure input can reproduce a family $\{p(\cdot\mid c)\}$ exactly only if its reference distribution $|\psi(z)|^2$, the $X$-basis distribution of its input, is compatible with every member. (ii) For every $n\geq2$ and every fixed IQP unitary, there are families with no common compatible $X$-basis distribution, which by (i) no fixed input reproduces, but which condition-dependent inputs with an efficient sampling-and-ratio interface reproduce exactly.
\end{proposition}

\noindent Appendix~\ref{sec:appendix_two_qubit_example} proves both parts. Part (i) holds for any generator set, and its condition is also sufficient when the diagonal layer contains all $2^n$ Pauli-$Z$ strings, a layer with exponentially many generators that is no longer an IQP circuit. Part (ii) uses a two-member family consisting of a deterministic outcome and a uniform distribution over three strings; for $n=2$, the latter is the distribution with which Kurkin \textit{et al.} showed that two-qubit IQP Born machines are not universal \cite[Lemma~11]{kurkinUniversalityKernelAdaptive2025}. Since the IQP unitary cannot change the $X$-basis distribution, the separation comes from the input, independent of the diagonal layer.

We call an encoder \emph{diagonal-equivalent} if its reference magnitudes $|\psi_c(z)|$ do not depend on the condition. Then $\ket{\psi_c}=\Lambda_c\ket{\psi}$ for a fixed state $\ket\psi$ and a condition-dependent diagonal unitary $\Lambda_c$, and the model is diagonal conditioning on a fixed input with the diagonal layer $D(\boldsymbol\theta)\Lambda_c$. Hence, only encoders that are not diagonal-equivalent can reach families beyond diagonal conditioning. Moreover, if the magnitudes are uniform at every condition, then $\ket{\psi_c}=\Lambda_c\ket{+}^{\otimes n}$, and the model is a standard IQP Born machine on $\ket0^{\otimes n}$ with diagonal layer $D(\boldsymbol\theta)\Lambda_c$. When $\Lambda_c=\exp\bigl(i\sum_{g\in\mathcal G}\lambda_g(c)\,g\bigr)$ is built from the ansatz generators, it merges with the learned diagonal layer,
\begin{equation}
D(\boldsymbol\theta)\,\Lambda_c=D\bigl(\boldsymbol\theta+\boldsymbol\lambda(c)\bigr),
    \label{eq:absorption}
\end{equation}
and the encoder reduces to a condition-dependent shift of the parameters.

Whether an encoder is diagonal-equivalent cannot be read off the gates that prepare the input, because the classification concerns the reference state, which includes the initial Hadamard layer of $U_{\boldsymbol\theta}$. The input $R_X(t)\ket{0}$, prepared by a non-diagonal gate, has $t$-dependent magnitudes, but the Hadamard layer turns them into phases. Then, its reference state $R_Z(t)\ket{+}$ has uniform magnitudes, and Eq.~\eqref{eq:absorption} absorbs it as $\theta_i\to\theta_i-t/2$. Conversely, the input $R_Z(t)\ket{+}$, prepared by a diagonal gate, is not diagonal-equivalent, because its reference state $R_X(t)\ket{0}$ has magnitudes that depend on $t$. This classification applies to any input; classical training additionally requires the sampling-and-ratio interface of Sec.~\ref{sec:model-access}, which every encoder in this work provides.

For the input $\ket{0}^{\otimes n}$, part (i) recovers the known non-universality of IQP Born machines \cite{recio-armengolTrainClassicalDeploy2026,kurkinUniversalityKernelAdaptive2025}. Kurkin \textit{et al.} restored universality by tracing out $n+1$ hidden qubits, at the cost of exponentially many phases \cite{kurkinUniversalityKernelAdaptive2025}. Hidden qubits also lift the obstruction of part (i): with a single one, diagonal conditioning reproduces every two-qubit distribution \cite[Lemma~12]{kurkinUniversalityKernelAdaptive2025}, and hence both members of the family of part (ii), which involve only two qubits. Input conditioning reaches such families without additional qubits, which suggests a trade-off between the number of hidden qubits and the complexity of the circuit that prepares the input state.

\subsection{Symmetries imposed by the encoder}
\label{sec:input-symmetry}

Many target families have symmetries known in advance, and models that respect them by construction need not learn them \cite{meyerExploitingSymmetry2023}. Encoding a symmetry in the input state imposes it on every member of the family while leaving the IQP unitary unrestricted.

For a bit mask $h\in\{0,1\}^n$, the bit-flip operator $X_h=\bigotimes_iX_i^{h_i}$ commutes with every IQP unitary, since $H^{\otimes n}X_hH^{\otimes n}=Z_h$ commutes with the diagonal layer. If the input is an eigenstate of $X_h$ at every condition, so is the output, whose amplitudes at $x$ and $x\oplus h$ differ at most by a sign, and hence $q_{\boldsymbol\theta}(x\mid c)=q_{\boldsymbol\theta}(x\oplus h\mid c)$ for every parameter vector and condition. Recio-Armengol \textit{et al.} introduced this construction for unconditional models, using GHZ inputs to impose global bit-flip symmetry \cite[Sec.~6]{recio-armengolTrainClassicalDeploy2026}.

Equivalently, the reference state is an eigenstate of $Z_h$, that is, all strings in its support have the same parity $h\cdot z\bmod2$, and the condition may change the amplitudes within that sector freely. The estimator is unaffected, since samples never leave the sector. In particular, even-parity reference states give $q_{\boldsymbol\theta}(x\mid c)=q_{\boldsymbol\theta}(\bar x\mid c)$, with $\bar x$ the bitwise complement of $x$; the parity encoders of Sec.~\ref{sec:cw_main} use this to impose global spin-flip symmetry on the Curie--Weiss models at every temperature.

\section{Conditional MMD objective}
\label{sec:cond_training}

We train conditional IQP Born machines by minimizing the squared maximum mean discrepancy (MMD) across training conditions. With a Gaussian kernel, this objective depends on the model only through its Pauli-$Z$ moments, which can be estimated classically as described in Sec.~\ref{sec:ratio-estimator}. For distributions $p$ and $q$ on $\{0,1\}^n$ and a positive-definite kernel $k$, the population squared MMD is \cite{grettonKernelTwoSample2012}
\begin{equation}
\begin{aligned}
    \mmd^2(p,q)
    ={}&\E_{x,x'\sim p}[k(x,x')]
    -2\E_{x\sim p,\,y\sim q}[k(x,y)] \\
    &+\E_{y,y'\sim q}[k(y,y')],
\end{aligned}
\label{eq:mmd-population}
\end{equation}
where the samples in each expectation are drawn independently.

For binary strings, $\|x-y\|_2^2$ equals the Hamming distance $d_H(x,y)$, and the Gaussian kernel with bandwidth $\sigma>0$ becomes $k_\sigma(x,y)=\exp[-d_H(x,y)/(2\sigma^2)]$. It depends only on the number of differing bits and treats all bits equally.

Let $\mathcal C_{\mathrm{train}}$ denote the set of training conditions and $\Sigma$ the set of kernel bandwidths. We define the condition-balanced objective as
\begin{equation}
    \mathcal L(\boldsymbol{\theta})
    =
    \frac{1}{|\mathcal C_{\mathrm{train}}|}
    \sum_{c\in\mathcal C_{\mathrm{train}}}
    \frac{1}{|\Sigma|}
    \sum_{\sigma\in\Sigma}
    \mmd_\sigma^2
    \!\left(
        p(\cdot\mid c),
        q_{\boldsymbol{\theta}}(\cdot\mid c)
    \right).
    \label{eq:conditional-objective}
\end{equation}
Here, $\mmd_\sigma^2$ denotes the MMD computed with kernel $k_\sigma$. Each discrepancy compares the target and the model at the same condition, and the objective contains no term that couples different conditions; learning across conditions enters only through the shared parameters $\boldsymbol\theta$.

To express the objective through Pauli-$Z$ moments, we apply the expansion of the Gaussian kernel in binary characters used in earlier work on QCBM and classical IQP training \cite{rudolphTrainabilityBarriersOpportunities2024,recio-armengolTrainClassicalDeploy2026} separately at each condition. Let $\pi_\sigma$ be the law of a random mask $a\in\{0,1\}^n$ with independent coordinates
\begin{equation}
    a_i\sim\operatorname{Bernoulli}(p_\sigma),
    \qquad
    p_\sigma=
    \frac{1-e^{-1/(2\sigma^2)}}{2}.
    \label{eq:mmd-mask-law}
\end{equation}
The Gaussian kernel is then the average of the characters $\chi_a$ of Eq.~\eqref{eq:iqp-phase-function} over this law,
\begin{equation}
    k_\sigma(x,y)=\E_{a\sim\pi_\sigma}\!\left[\chi_a(x)\chi_a(y)\right],
    \label{eq:kernel-characters}
\end{equation}
since each bit in which $x$ and $y$ differ contributes a factor $1-2p_\sigma=e^{-1/(2\sigma^2)}$. With the Pauli-$Z$ moments of Sec.~\ref{sec:model-access}, $\mu_q(a\mid c)=\E_{x\sim q_{\boldsymbol\theta}(\cdot\mid c)}[\chi_a(x)]$, and the target moments $\mu_p(a\mid c)$ defined in the same way, the squared MMD becomes
\begin{equation}
    \mmd_\sigma^2
    \!\left(
        p(\cdot\mid c),
        q_{\boldsymbol{\theta}}(\cdot\mid c)
    \right)
    =
    \E_a
    \!\left[
        \left(
            \mu_p(a\mid c)-\mu_q(a\mid c)
        \right)^2
    \right].
    \label{eq:mmd-character-moments}
\end{equation}
The squared MMD is thus a weighted squared distance between the Walsh--Fourier spectra of the two distributions, with weights $\pi_\sigma(a)$. Since $0<p_\sigma<1/2$, every mask has positive weight, and the MMD vanishes if and only if $p(\cdot\mid c)=q_{\boldsymbol\theta}(\cdot\mid c)$. The weights of high-order masks can, however, be exponentially small, so we also evaluate the TVD and physical observables.

The bandwidth sets which correlation orders the objective emphasizes. The Hamming weight $|a|$, which is the order of $Z_a$, follows $\operatorname{Binomial}(n,p_\sigma)$, and we call its mean $\bar\ell=np_\sigma$ the average bodyness, adapting the term of Rudolph \textit{et al.} \cite{rudolphTrainabilityBarriersOpportunities2024}. Larger bandwidths emphasize lower orders, and $\bar\ell\to n/2$ as $\sigma\to0$; averaging over several bandwidths in Eq.~\eqref{eq:conditional-objective} combines these regimes. Emphasizing low orders also mitigates barren plateaus \cite{mccleanBarrenPlateausQuantum2018,laroccaBarrenPlateausVariational2025,cerezoCostFunctionDependent2021}, at the cost of sensitivity to higher-order differences \cite{rudolphTrainabilityBarriersOpportunities2024}; dedicated mitigation strategies for IQP Born machines \cite{shenCharacterizingTrainability2026,lerchIQPInitialization2026,kurkinUniversalityKernelAdaptive2025} are outside our scope.

\subsection{Estimating the objective}
\label{sec:unbiased-conditional-mmd}

Each optimizer update draws a training condition $c$ and a bandwidth $\sigma$ uniformly, a minibatch of $B$ target samples $x_b$ with replacement from the training pool at $c$, $S$ independent reference samples $z_s$ from $|\psi_c|^2$, and $M$ masks, with $B,S\geq2$. All masks are evaluated on the same two batches, which correlates the per-mask estimates but leaves each unbiased for the MMD between the model and the empirical distribution of the pool. For each mask $a$, the estimator of Recio-Armengol \textit{et al.} \cite[Prop.~3]{recio-armengolTrainClassicalDeploy2026}, extended to condition-dependent reference states, is
\begin{equation}
\begin{aligned}
    \widehat{\delta_c^2}(a)
    ={}&
    \frac{1}{S(S-1)}
    \sum_{s\neq t}Y_sY_t
    -2\overline Y\,\overline T \\
    &+
    \frac{1}{B(B-1)}
    \sum_{b\neq b'}T_bT_{b'} ,
\end{aligned}
\label{eq:conditional-mmd-ustatistic}
\end{equation}
where $T_b=\chi_a(x_b)$ and $Y_s=Y_{c,a}(z_s;\boldsymbol\theta)$ are single-sample estimates of the target and model moments, and bars denote sample means. The estimator is unbiased for $\bigl(\hat\mu_p(a\mid c)-\mu_q(a\mid c)\bigr)^2$, where $\hat\mu_p$ is the moment of the training pool, because its pair sums exclude self-products \cite{grettonKernelTwoSample2012} and distinct samples are independent; it can take negative values. Its gradient,
\begin{equation}
    \partial_{\theta_g}
    \widehat{\delta_c^2}(a)
    =
    \frac{2}{S(S-1)}
    \sum_{s\neq t}Y_sY'_t
    -
    2\overline T\,\overline{Y'},
    \label{eq:conditional-mmd-gradient}
\end{equation}
is also unbiased, with $Y'_s=Y'_{g,c,a}(z_s;\boldsymbol\theta)$ the single-sample gradient estimates of Eq.~\eqref{eq:ratio-sample} and $\overline{Y'}$ their mean. Each mask and gradient component costs $O(S+B)$ operations once the ratios and phase differences are available, since the pair sums reduce to sums and sums of squares. Since the identity mask contributes nothing, we draw the $M$ masks from $\pi_\sigma(a\mid a\neq0)$ and multiply their average by $Q_\sigma^+=1-(1-p_\sigma)^n$,
\begin{equation}
    \widehat{\mmd}_{\sigma}^2(c)
    =
    Q_\sigma^+
    \frac{1}{M}
    \sum_{m=1}^{M}
    \widehat{\delta_c^2}(a_m),
    \label{eq:sampled-mmd-estimator}
\end{equation}
which is unbiased for Eq.~\eqref{eq:mmd-character-moments} with $p(\cdot\mid c)$ replaced by the empirical distribution of the pool. We quote $\bar\ell=np_\sigma$ throughout, although the sampled masks have mean order $np_\sigma/Q_\sigma^+$. The variances of these estimators are bounded independently of the number of qubits (Appendix~\ref{sec:estimator-variance}).

\section{Target distribution families}
\label{sec:targets}

\begin{figure*}[!htb]
\centering
\includegraphics[width=\textwidth]{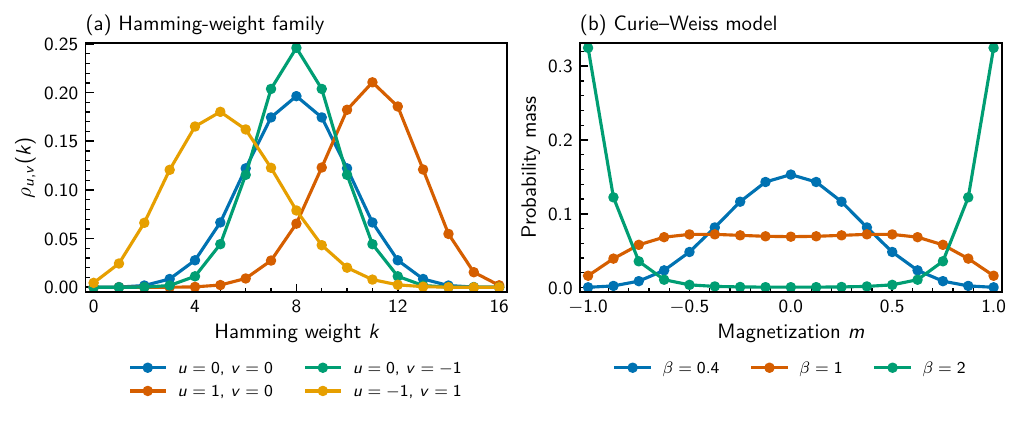}
\caption{Target families. (a) Hamming-weight profiles $\rho_{u,v}(k)$ of the two-dimensional family: the reference $(u,v)=(0,0)$, a shift of the mean, a change of the width, and both combined. (b) Magnetization profiles of the sixteen-spin Curie--Weiss model across the crossover.}
\label{fig:target-families}
\end{figure*}

Both target families assign equal probability to bitstrings with the same Hamming weight and are invariant under permutations of the bits. This symmetry matches the Hamming-distance kernel of the MMD objective. The two families play complementary roles. The first has two independent conditioning coordinates and tests whether the model generalizes to their combinations. In the second, the Curie--Weiss model, the magnetization profile changes from unimodal to bimodal as the temperature decreases, which requires a single model to capture a qualitative change in the shape of the distribution across conditions. Its observables remain classically tractable at any size, which lets us validate models trained with up to $2048$ qubits. Figure~\ref{fig:target-families} shows representative members of both families.

\subsection{Hamming-weight family with tunable mean and width}
\label{sec:target-gaussian}

The first family is indexed by a two-dimensional condition $c=(u,v)$, where $u$ controls the location and $v$ the width of a Hamming-weight distribution. We deform the binomial Hamming-weight profile by a quadratic exponential factor. For $u,v\in[-1,1]$, the target Hamming-weight profile is
\begin{equation}
    \rho_{u,v}(k)\propto\binom{16}{k}
    \exp\!\left[-\frac{(k-\bar k_u)^2}{2s_v^2}
    +\frac{(k-8)^2}{2s_0^2}\right],
\end{equation}
normalized over $k\in\{0,\ldots,16\}$, with $\bar k_u=8+3u$, $s_v=2(1+0.2v)$ and the binomial width $s_0=2$. The second term cancels the Gaussian shape of the binomial coefficient to leading order, so the profile is close to a Gaussian with center $\bar k_u$ and width $s_v$; $u$ and $v$ thus control its location and width without setting its moments exactly. We then assign equal probability to bitstrings with the same Hamming weight $|x|=\sum_ix_i$,
\begin{equation}
    p(x\mid u,v)=\frac{\rho_{u,v}(|x|)}{\binom{16}{|x|}}.
    \label{eq:gaussian-two-dimensional-target}
\end{equation}
Along $v=0$, the two quadratic terms in the exponent cancel, leaving a term linear in $k$, and the targets reduce to product distributions with binomial Hamming-weight profiles. At the origin, the target is uniform over bitstrings, and its Hamming-weight profile is binomial.

\subsection{Curie--Weiss model}
\label{sec:target-cw}

The second family is the zero-field Curie--Weiss spin model \cite{ellisEntropyLargeDeviations2006}, conditioned on the inverse temperature $\beta$. For $n$ spins, we represent each configuration by a bitstring $x\in\{0,1\}^n$, with $s_i=2x_i-1$. The magnetization per spin is then $m(x)=n^{-1}\sum_i s_i=(2|x|-n)/n$. With unit interaction strength, the energy is $H(x)=-n\,m(x)^2/2$, and the target is the Gibbs distribution at inverse temperature $\beta$,
\begin{equation}
    p(x\mid\beta)=\frac{e^{-\beta H(x)}}{\mathcal Z_\beta}=\frac{1}{\mathcal Z_\beta}
    \exp\!\left[\frac{\beta}{2n}(2|x|-n)^2\right],
    \label{eq:cw_target}
\end{equation}
where $\mathcal Z_\beta$ is the normalization constant. In addition to permutation invariance, the target exhibits global spin-flip symmetry. The crossover from the disordered to the ordered regime occurs near $\beta=1$ and becomes sharper as $n$ increases.

We characterize this crossover using two observables. The mean absolute magnetization $\langle|m|\rangle$ measures the magnetic order; we use the absolute value because spin-flip symmetry makes the signed mean vanish. The per-spin energy-fluctuation statistic
\begin{equation}
    c_{\mathrm{fluc}}(\beta)=\frac{\beta^2}{n}
    \bigl(\langle H^2\rangle-\langle H\rangle^2\bigr)
    \label{eq:cw_heat_fluctuation}
\end{equation}
equals the heat capacity per spin for the Gibbs target, and its peak indicates the crossover. Since the target depends on $x$ only through $|x|$, its observables follow exactly from sums over $n+1$ Hamming weights at any size, which allows trained models to be validated far beyond exact simulation (Sec.~\ref{sec:results-scaling}).

\section{Numerical experiments}
\label{sec:numerical-experiments}
\label{sec:results}

We test the framework in four experiments. We train all IQP models by minimizing the sampled squared MMD of Sec.~\ref{sec:cond_training}. Within each experiment, all models, including the baselines, are selected by the same hyperparameter-optimization protocol on development conditions and evaluated on separate conditions outside the training set. The per-$\beta$ circuits of Sec.~\ref{sec:cw_main} are the exception, since they serve as a reference trained directly at the evaluation $\beta$ (Appendix~\ref{sec:appendix_protocols}). At sixteen qubits, our accuracy metric is the exact total variation distance,
\begin{equation}
    \operatorname{TVD}(p,q)
    =\frac{1}{2}\sum_{x\in\{0,1\}^n}|p(x)-q(x)|,
    \label{eq:full-distribution-tvd}
\end{equation}
which we compute from exact circuit probabilities, average over evaluation conditions within a fit, and report as the mean and standard deviation across independent fits, which differ in their random seed. For Hamming-weight profiles, we also use the coarser profile TVD, $\operatorname{TVD}_K$, the TVD between the Hamming-weight distributions $\rho_p$ and $\rho_q$ of the two distributions, which never exceeds the full TVD.

\subsection{Generalization to unseen condition combinations}
\label{sec:gaussian-two-dimensional}
\label{sec:results-gaussian}

The first experiment tests whether a conditional model trained with one condition coordinate varied at a time generalizes to combinations of both coordinates, on the Hamming-weight family of Sec.~\ref{sec:target-gaussian} with sixteen qubits.

\begin{figure*}[!htb]
\centering
\includegraphics{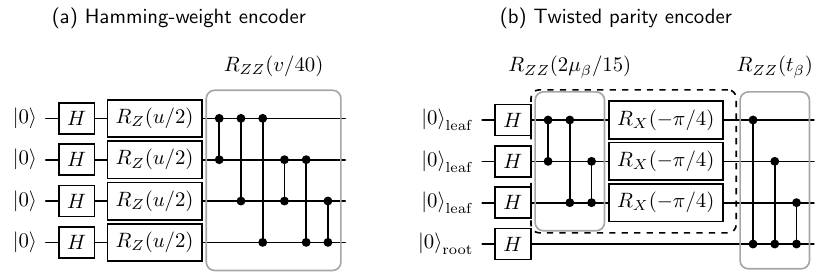}
\caption{Encoder circuits, drawn for four qubits with the angles of the sixteen-qubit models, where $R_Z(\varphi)=e^{-i\varphi Z/2}$ and connected dots denote $R_{ZZ}(\varphi)=e^{-i\varphi Z\otimes Z/2}$ with the angle shown above each group. (a) Hamming-weight encoder of Eq.~\eqref{eq:gaussian-two-dimensional-input}. (b) Twisted parity encoder of Eq.~\eqref{eq:cw_twisted_input}; without the dashed block, which contains the twist and the rotation, it reduces to the parity encoder of Eq.~\eqref{eq:cw_parity_input} with angle $t_\beta$.}
\label{fig:encoder-circuits}
\end{figure*}

We encode the two coordinates in the phases of the uniform superposition (Fig.~\ref{fig:encoder-circuits}(a)),
\begin{equation}
    \ket{\phi(u,v)}
    =e^{-i(u/2)J_z}e^{-i(v/40)J_z^2}
    \ket{+}^{\otimes16},
    \label{eq:gaussian-two-dimensional-input}
\end{equation}
where $J_z=\tfrac12\sum_{j=1}^{16}Z_j$ is the collective spin. Since $J_z\ket{x}=(8-|x|)\ket{x}$, the two rotations give phases linear and quadratic in the Hamming weight $|x|$, matched to the location and the width of the target. Both are diagonal and leave the input probabilities uniform. The reference magnitudes nonetheless depend on $(u,v)$, so the encoder is not diagonal-equivalent (Sec.~\ref{sec:gauge}), and the IQP unitary must convert the encoded phases into the target probabilities. At the origin, the input is $\ket{+}^{\otimes16}$, which the IQP unitary preserves up to a global phase, and the model matches the uniform target by construction. The amplitudes depend only on the Hamming weight, which makes sampling and amplitude ratios efficient (Appendix~\ref{sec:gaussian-two-dimensional-protocol}).

We use an IQP unitary with $40$ trainable phases, one on each qubit and one on each of the $24$ nearest-neighbor pairs of a $4\times4$ square lattice. We train at nine conditions that vary one coordinate at a time with the other fixed at zero, which span the diamond $|u|+|v|\leq1$. We select models on eight further development conditions on the axes and report their accuracy at sixteen combinations of both coordinates, inside and outside this convex hull, which enter neither training nor selection.

The diagonal-conditioning baseline applies the same lattice circuit to the fixed input $\ket{0}^{\otimes16}$ and makes its $40$ phases an affine function of the condition, $\boldsymbol\theta(u,v)=\mathbf w_0+u\,\mathbf w_u+v\,\mathbf w_v$, with $120$ parameters in total. With reference state $\ket{+}^{\otimes16}$ at every condition, it is diagonal-equivalent (Sec.~\ref{sec:gauge}). We include no term in $uv$: every training condition has $u=0$ or $v=0$, so such a term would vanish throughout training and could not be learned.

\begin{figure*}[!t]
 \centering
 \includegraphics[width=0.85\textwidth]{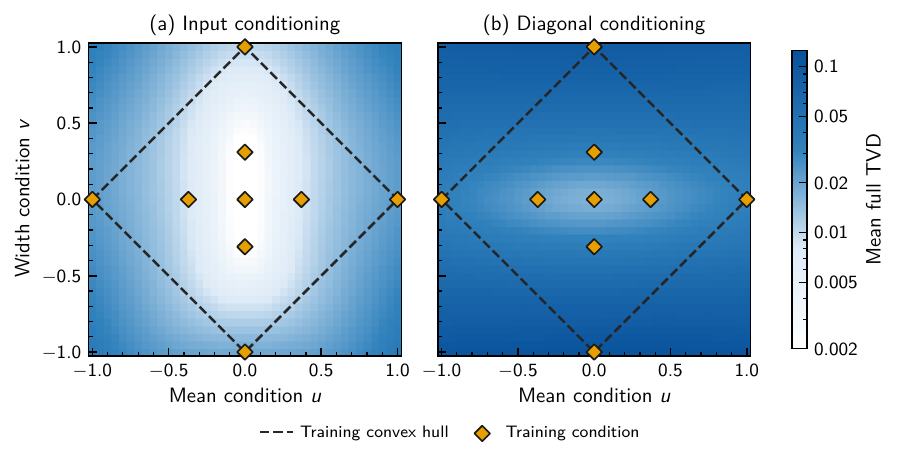}
 \caption{Joint conditional generation of the Hamming-weight family, (a) with input conditioning and (b) with diagonal conditioning of the same lattice circuit acting on $\ket{0}^{\otimes16}$. Colors show the mean full-distribution TVD across ten seeds on a $41\times41$ grid of conditions, on a shared logarithmic scale. Diamonds mark the nine training conditions, which lie on the two axes, and the dashed line marks their convex hull. At the origin, the target and the input-conditioned model are uniform by construction.}
 \label{fig:gaussian-two-dimensional}
\end{figure*}

Figure~\ref{fig:gaussian-two-dimensional}(a) shows that the conditional model reproduces the target family across the evaluated condition square, although it was trained only on the two axes. At the sixteen mixed conditions, its TVD is $0.0216\pm0.0017$ (Table~\ref{tab:gaussian-two-dimensional} in Appendix~\ref{sec:gaussian-two-dimensional-protocol}), with $0.0085\pm0.0012$ at the four conditions inside the training hull and $0.0259\pm0.0018$ at the twelve outside it. Across the condition square, its error correlates more strongly with each target's TVD from the uniform distribution ($0.91$) than with its distance to the nearest training condition ($0.72$) (Appendix~\ref{sec:gaussian-two-dimensional-protocol}).

Diagonal conditioning reaches a TVD of $0.069\pm0.002$ at the mixed conditions under the same protocol, about three times that of input conditioning. The gap is already present at the training conditions other than the origin, with mean TVD $0.048\pm0.002$ against $0.0117\pm0.0011$, which rules out a purely extrapolation effect. It comes mainly from the width coordinate. The diagonally conditioned model follows the shift of the mean closely but changes the width by only $3$--$4\%$ of the target change, so its error is largest at the corners of the square, where it reaches $0.085$--$0.109$ against about $0.035$ for input conditioning (Fig.~\ref{fig:gaussian-two-dimensional}). More flexible diagonal variants, with quadratic condition dependence or all-to-all connectivity, did not improve on the affine map under this protocol (Appendix~\ref{sec:gaussian-two-dimensional-protocol}). With three times as many parameters, diagonal conditioning thus fails to learn a condition dependence that input conditioning captures.

\subsection{Conditional modeling of the Curie--Weiss model}
\label{sec:cw_main}
\label{sec:results-cw}

The second experiment tests whether a single sixteen-qubit model learns how the Curie--Weiss distribution of Sec.~\ref{sec:target-cw} changes with temperature, and how much the encoder matters.

Both encoders we use treat one qubit, which we call the root (qubit $16$), differently from the other fifteen, which we call the leaves and treat identically; $J_z=\tfrac12\sum_{i=1}^{15}Z_i$ is the collective spin of the leaves. The parity encoder applies $ZZ$ rotations between the root and every leaf,
\begin{equation}
 \ket{\phi_{\mathrm P}(\beta)}=
 \exp\!\left[-\frac{i t}{2}\sum_{i=1}^{15}Z_iZ_{16}\right]
 \ket{+}^{\otimes16},
 \label{eq:cw_parity_input}
\end{equation}
with angle $t$. Because its reference state is supported on even-parity strings, the output is invariant under a global spin flip (Sec.~\ref{sec:input-symmetry}). Its input probabilities are uniform, but it is not diagonal-equivalent, because its reference magnitudes depend on $\beta$. Every Hamming weight enters with its binomial weight, and the IQP unitary must produce the entire temperature dependence of the profile from phases.

The twisted parity encoder adds two leaf operations before the root--leaf rotations (Fig.~\ref{fig:encoder-circuits}(b)),
\begin{equation}
\begin{split}
 \ket{\phi_{\mathrm{TP}}(\beta)}={}&
 \exp\!\left[-\frac{i t_\beta}{2}\sum_{i=1}^{15}Z_iZ_{16}\right]\\
 &\times\bigl[R_X(-\tfrac{\pi}{4})^{\otimes15}e^{-i(2\mu_\beta/15)J_z^2}\ket{+}^{\otimes15}\bigr]\otimes\ket{+}.
\end{split}
 \label{eq:cw_twisted_input}
\end{equation}
Together, the two leaf operations form the one-axis-twisting protocol for spin squeezing \cite{kitagawaSqueezedSpinStates1993}, and the twist generator $J_z^2$ is proportional to the Curie--Weiss energy of the leaves. The twisted parity encoder thus builds the form of the target interaction into the input, whereas the parity encoder imposes only the spin-flip symmetry. The twist narrows the transverse spread of the collective spin, initially along $x$, and the rotation about $x$ turns this narrowed direction toward $z$, so the leaf Hamming-weight distribution becomes narrower than binomial and the input probabilities depend on temperature. We fix the rotation angle at $\pi/4$, close to the angle that narrows this distribution most for our twist strengths (Appendix~\ref{sec:appendix_twisted}). Since $ZZ$ rotations are invariant under the global flip and $R_X$ commutes with $X$, every layer commutes with $X^{\otimes16}$, and both encoders impose $q_{\boldsymbol\theta}(x\mid\beta)=q_{\boldsymbol\theta}(\bar x\mid\beta)$ for every parameter vector. Every layer is also invariant under permutations of the leaves, which makes sampling and amplitude ratios efficient (Appendix~\ref{sec:appendix_cw}).

The angles of the twisted parity encoder follow the fixed schedule
\begin{equation}
 \sin t_\beta=\tanh(\eta\beta^2),\qquad \mu_\beta=\kappa\beta\cos t_\beta ,
 \label{eq:cw_twisted_schedule}
\end{equation}
with $t_\beta\in[0,\pi/2)$. The root--leaf angle rises from $0$ and saturates at $\pi/2$. This mirrors the target, which concentrates on the two fully magnetized states as $\beta$ grows, and keeps inputs beyond the largest training temperature close to those seen in training. The twist strength $\mu_\beta$ vanishes at $\beta=0$ and as $\beta\to\infty$, so the encoder twists mainly around the crossover. With $\mu_\beta=0$, the rotation acts on $\ket{+}^{\otimes15}$ as a global phase, and the encoder reduces to the parity encoder with angle $t_\beta$. At $\beta=0$, the input is $\ket{+}^{\otimes16}$ and the model again matches the uniform target by construction. The constant $\kappa=0.4$ and the rotation angle are fixed, and $\eta$ is selected by the hyperparameter search (Appendix~\ref{sec:cw_hpo}). None of the encoder parameters are updated by the optimizer.

We use an IQP unitary with all sixteen one-body and $120$ two-body generators, giving $136$ trainable phases per circuit. We train the conditional models at $\beta\in\{0.1,0.7,1.3,1.95\}$ and evaluate them at $\beta\in\{0.4,1,2\}$, the last slightly beyond the largest training temperature, and on a separate grid of $101$ temperatures in $[0,2]$, computing the TVD and the two observables of Sec.~\ref{sec:target-cw}.

We write C-IQP for a conditional IQP model that shares one circuit across conditions and add its encoder in parentheses; our model is C-IQP (TP), with TP standing for twisted parity. Table~\ref{tab:cw_models} lists the IQP-based alternatives. C-IQP (P) uses the parity encoder with the linear angle $t=\lambda\beta$. The per-$\beta$ circuits are trained directly at the evaluation $\beta$, so they measure how well a single circuit fits each target rather than generalization, and test whether sharing one circuit across temperatures costs accuracy.

The main baseline is diagonal conditioning, the approach of previous conditional IQP models, which we call diagonally conditioned IQP (D-IQP). Whereas previous approaches fit the phases separately at each condition and interpolate between them \cite{liuCorrelationComplexity2026,liuPhysicsInformedLatent2026,zouConQuER2025}, D-IQP is trained jointly across temperatures like our model. It uses one circuit with the same $136$ generators, whose phases are a cubic polynomial in the condition, $\boldsymbol\theta(\beta)=\sum_{d=0}^{3}T_d(\beta-1)\,\mathbf w_d$, with Chebyshev polynomials $T_d$. Since four parameter vectors determine a unique cubic, it can represent any set of per-temperature fits at our four training temperatures, interpolated by a cubic. Its fixed input, the parity encoder at $t=\pi/2$, imposes the same spin-flip symmetry.

\begin{table*}[t]
\centering
\caption{Sixteen-qubit Curie--Weiss models. All circuits use the same $136$ one- and two-body generators. The shared models use one circuit for all $\beta$, and the per-$\beta$ models one circuit per evaluation $\beta$. Parameters count all phases or weights updated by the optimizer.}
\label{tab:cw_models}
\small
\begin{tabular}{@{}llll@{}}
\toprule
Model & Input and conditioning & Parameters & Difference from C-IQP (TP) \\
\midrule
C-IQP (TP) & \parbox[t]{0.24\textwidth}{\raggedright twisted parity, Eq.~\eqref{eq:cw_twisted_schedule}\strut} & $136$ & \parbox[t]{0.38\textwidth}{\raggedright (proposed model)\strut} \\\addlinespace[3pt]
C-IQP (P) & \parbox[t]{0.24\textwidth}{\raggedright parity, $t=\lambda\beta$\strut} & $136$ & \parbox[t]{0.38\textwidth}{\raggedright no twist, and a linear angle $t=\lambda\beta$ instead of the schedule of Eq.~\eqref{eq:cw_twisted_schedule}\strut} \\\addlinespace[3pt]
C-IQP (TP), $\kappa=0$ & \parbox[t]{0.24\textwidth}{\raggedright parity, Eq.~\eqref{eq:cw_twisted_schedule}\strut} & $136$ & \parbox[t]{0.38\textwidth}{\raggedright twist removed (ablation, Appendix~\ref{sec:cw_product_ablation})\strut} \\\addlinespace[3pt]
D-IQP & \parbox[t]{0.24\textwidth}{\raggedright parity at $t=\pi/2$, cubic phases\strut} & $544$ & \parbox[t]{0.38\textwidth}{\raggedright fixed input, with the condition in the diagonal-layer phases\strut} \\\addlinespace[3pt]
Per-$\beta$ IQP (TP) & \parbox[t]{0.24\textwidth}{\raggedright twisted parity at each $\beta$\strut} & $3\times136$ & \parbox[t]{0.38\textwidth}{\raggedright same encoder, with one circuit trained at each evaluation $\beta$\strut} \\\addlinespace[3pt]
Per-$\beta$ IQP & \parbox[t]{0.24\textwidth}{\raggedright $\ket{0}^{\otimes16}$\strut} & $3\times136$ & \parbox[t]{0.38\textwidth}{\raggedright one circuit per evaluation $\beta$, on the standard input without the spin-flip symmetry\strut} \\
\bottomrule
\end{tabular}
\end{table*}

\begin{figure*}[!t]
 \centering
 \includegraphics[width=\linewidth]{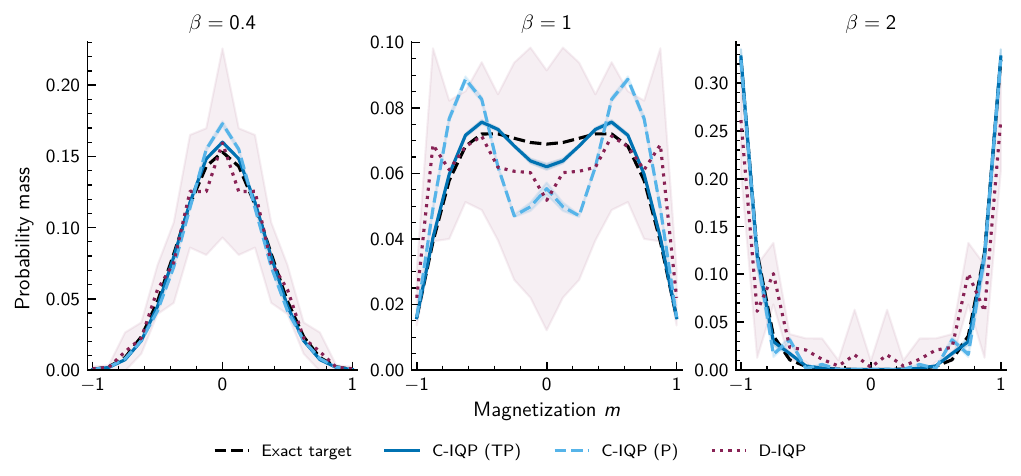}
 \caption{Magnetization profiles of the sixteen-qubit Curie--Weiss models at the evaluation $\beta$, for the conditional models with the twisted parity encoder, C-IQP (TP), and with the parity encoder, C-IQP (P), and for diagonal conditioning, D-IQP. Magnetization per spin is $m=(2k-16)/16$, where $k$ is Hamming weight. Colored curves show exact probability masses averaged across ten independent fits, and shading shows one standard deviation above and below the mean. Dashed black curves show the targets. The D-IQP band is drawn lighter for readability. It is wide because most D-IQP fits alternate between high and low probability on neighboring Hamming weights, each with a different pattern, which the mean smooths out. The displayed values of $\beta$ are excluded from training and model selection. Panels use separate vertical scales.}
 \label{fig:curie_weiss_phase_flip}
\end{figure*}

\begin{figure*}[t]
 \centering
 \includegraphics[width=\linewidth]{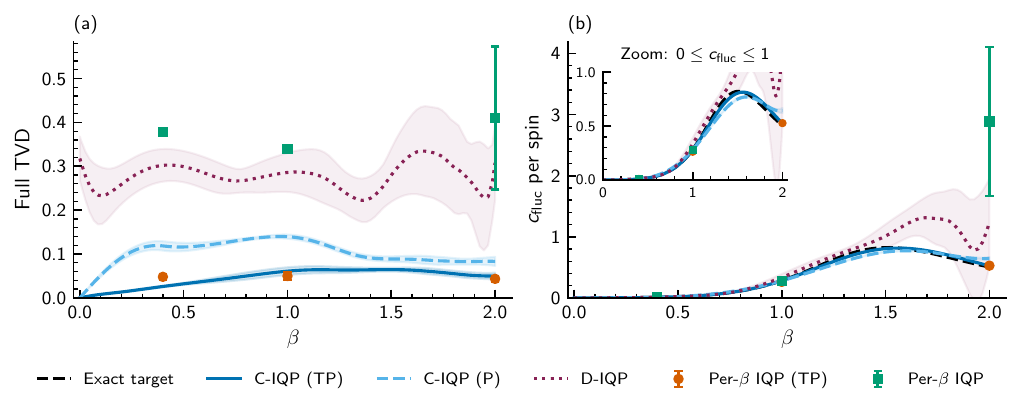}
 \caption{Sixteen-qubit Curie--Weiss models across temperature: (a) exact full-distribution TVD and (b) per-spin energy fluctuations $c_{\mathrm{fluc}}$ on the $101$-point evaluation grid. Curves show the models of Fig.~\ref{fig:curie_weiss_phase_flip}, C-IQP (TP) (solid), C-IQP (P) (dashed), and D-IQP (dotted), with the target dashed black in (b); curves and bands show means and one standard deviation over ten fits. Markers show the per-$\beta$ IQP (TP) (circles) and per-$\beta$ IQP (squares) circuits, trained at the evaluation $\beta$. The inset in (b) zooms into the range of the target fluctuations.}
 \label{fig:cw_tvd_fluct}
\end{figure*}

We compare C-IQP (TP) with D-IQP, C-IQP (P), and the per-$\beta$ circuits. Figure~\ref{fig:curie_weiss_phase_flip} shows that C-IQP (TP) captures the change from a central magnetization peak in the hot regime ($\beta=0.4$) to two peaks at large absolute magnetization in the cold regime ($\beta=2$). Near the crossover at $\beta=1$, it follows the broad target profile, with only a shallow dip at $m=0$, whereas C-IQP (P) places too much probability at large $|m|$ and too little near $m=0$. D-IQP follows the envelope of the target profiles only roughly, distributing probability unevenly between neighboring Hamming weights. Much of its error lies within the Hamming-weight shells, among strings that a profile does not distinguish.

Figure~\ref{fig:cw_tvd_fluct}(a) compares the errors. Averaged over the evaluation grid, the TVD is $0.046\pm0.004$ for C-IQP (TP), $0.103\pm0.004$ for C-IQP (P), and $0.28\pm0.03$ for D-IQP. The twisted parity encoder lowers the mean TVD relative to the parity encoder at every nonzero grid point and in all ten seeds, which share training data and random streams between the two models. Its error grows through the hot regime and stays near its maximum between $\beta\approx1.1$ and $1.6$, whereas C-IQP (P) has its largest errors at the crossover. An ablation that removes only the twist raises the averaged TVD to $0.132\pm0.004$, above C-IQP (P), in all ten seeds, so the improvement comes from the twist (Appendix~\ref{sec:cw_product_ablation}). Diagonal conditioning, although its $544$ parameters can fit each training temperature separately, learns this target far less accurately than either conditional model under the same protocol. This already holds for C-IQP (P), whose encoder shares the spin-flip symmetry of the D-IQP input and adds only a linear angle, with a TVD $2.7$ times lower. The gain over diagonal conditioning therefore does not depend on the target-specific twist, which lowers the TVD by a further factor of two.

Averaged over the evaluation $\beta\in\{0.4,1,2\}$, sharing one circuit across temperatures produces no appreciable loss in TVD. The per-$\beta$ IQP (TP) circuits, trained directly at these temperatures with three times the total parameters and optimizer updates across the three-circuit family, reach a TVD of $0.047\pm0.005$, against $0.046\pm0.004$ for C-IQP (TP), which never sees these temperatures in training. The shared model is more accurate at $\beta=0.4$ and less accurate at $\beta=1$ and $2$. The per-$\beta$ IQP circuits, whose standard input imposes no spin-flip symmetry, have substantially higher TVD at every evaluation $\beta$ and vary strongly across fits at $\beta=2$.

Figure~\ref{fig:cw_tvd_fluct}(b) shows the energy fluctuations. C-IQP (TP) reproduces the fluctuation peak in both height and position, $0.815$ at $\beta=1.56$ against the target $0.823$ at $\beta=1.50$. Since $H=-nm^2/2$, the fluctuation statistic depends on the fourth moment of the magnetization and is therefore sensitive to the tails of the profile, where similar-looking profiles can differ. At $\beta=2$, for example, C-IQP (P) has $1.7$ times the full TVD of C-IQP (TP) ($0.083$ against $0.050$) but seven times its error in $c_{\mathrm{fluc}}$ normalized by the target peak ($0.17$ against $0.025$). Table~\ref{tab:cw_observable_errors} in Appendix~\ref{sec:cw_physical_observables} compares the observable errors of the main models at $\beta\in\{0.4,1,2\}$.

\subsection{Comparison with classical models}
\label{sec:cw_classical_setup}
\label{sec:cw_classical}

The third experiment tests whether conditional IQP Born machines are competitive with classical generative models on the sixteen-qubit Curie--Weiss task.

We train three classical conditional models: a restricted Boltzmann machine (RBM) with condition-dependent biases \cite{hintonTrainingProductsExperts2002,taylorModelingHumanMotion2007}, an autoregressive Transformer \cite{vaswaniAttentionAllYou2017}, and a conditional variational autoencoder (VAE) \cite{kingmaAutoEncodingVariationalBayes2014,sohnConditionalGenerative2015}. Each model receives $\beta$ as the conditioning variable, and we average the probabilities each model assigns to $x$ and $\bar x$, during both training and evaluation, to impose the same symmetry that the IQP encoders impose. The RBM and Transformer train on the exact likelihood and the VAE on its evidence lower bound, and parameter counts include all parameters learned in training, among them the VAE encoder. We use two sizes of each model: small models with a number of learned parameters comparable to C-IQP, and large models with between $950$ and $986$. All models follow the hyperparameter-optimization protocol of the IQP models, including the training temperatures and target samples (Appendix~\ref{sec:cw_classical_details}).

\begin{figure*}[t]
 \centering
 \includegraphics[width=\linewidth]{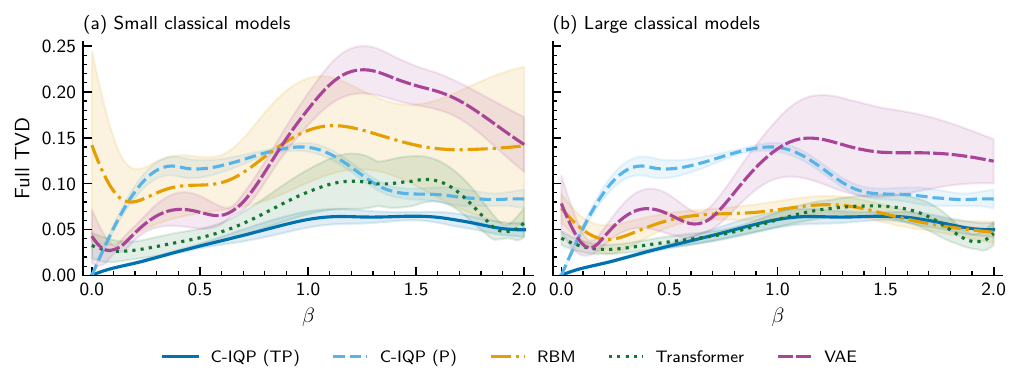}
 \caption{Exact full-distribution TVD of C-IQP (TP) and C-IQP (P), each with $136$ learned parameters, and of classical conditional models on the $101$-point Curie--Weiss evaluation grid: (a) small models with $134$ to $140$ and (b) large models with $950$ to $986$ learned parameters. Curves and bands show means and one standard deviation over ten fits.}
 \label{fig:cw_classical}
\end{figure*}

Figure~\ref{fig:cw_classical} shows their errors across the evaluation grid, and Table~\ref{tab:cw_classical} in Appendix~\ref{sec:cw_classical_details} lists the aggregate scores. At comparable learned-parameter counts, C-IQP (TP) achieves lower mean TVD than the three tested classical models, in every seed; the closest small model, the Transformer, reaches $0.068\pm0.015$ against $0.046\pm0.004$. Among the large models, with about seven times as many parameters, only the Transformer is comparable to C-IQP (TP). It reaches $0.052\pm0.005$, higher than C-IQP (TP) in eight of ten seeds, whereas the large RBM and VAE reach $0.061\pm0.007$ and $0.10\pm0.03$. It reproduces the mean absolute magnetization slightly more accurately, with an error of $0.006$ against $0.007$, but the normalized fluctuations less accurately, with $0.014$ against $0.010$ (Table~\ref{tab:cw_observable_errors}). With the parity encoder, C-IQP (P) falls behind the small Transformer and the large RBM and Transformer.

Three aspects of the training setup should be kept in mind when interpreting this comparison. The classical models train on likelihood-based objectives, exact for the RBM and Transformer at this size and an evidence lower bound for the VAE, whereas the C-IQP models minimize a sampled MMD estimate. Conversely, the twisted parity encoder applies an angle schedule chosen for this target, whereas the classical models learn their own dependence on $\beta$. Finally, the twisted parity encoder builds the form of the Curie--Weiss interaction into an input that is symmetric under permutations of fifteen of the sixteen qubits, whereas the classical models must learn this structure from data. Neither model class imposes the full permutation symmetry of the target.

\subsection{Large-scale training}
\label{sec:scaling}
\label{sec:results-scaling}

The fourth experiment scales training to $n\in\{64,128,256,512,1024,2048\}$ qubits, beyond the reach of exact simulation, and validates the models by comparing classically estimated observables with exact values (Sec.~\ref{sec:target-cw}).

These models use the parity encoder, which needs only a tree of nearest-neighbor CNOT gates, whereas the twist couples every pair of leaves (Sec.~\ref{sec:discussion}). Its angle rises with $\beta$ and saturates, for the same reason as in Sec.~\ref{sec:cw_main}, and its crossover can sharpen with $n$ through a steepness exponent selected from data (Appendix~\ref{sec:appendix_scaling}).

The IQP unitary has trainable phases only on each qubit and each nearest-neighbor pair of an open rectangular lattice. An $L_x\times L_y$ lattice thus has $3n-L_x-L_y$ parameters, against $n(n+1)/2$ with all-to-all connectivity. The trained models have between $176$ parameters at $64$ qubits and $6048$ at $2048$ qubits (Appendix~\ref{sec:appendix_scaling}), which keeps training efficient at these sizes. Since exact probabilities are unavailable at these sizes, we select hyperparameters and checkpoints by the classically estimated MMD at development temperatures, with searches at $64$ and $256$ qubits whose selected configuration is used at every size, and we validate the models through the mean squared magnetization $\langle m^2\rangle$ (Appendix~\ref{sec:appendix_scaling}).

\begin{figure*}[t]
    \centering
    \includegraphics[width=\linewidth]{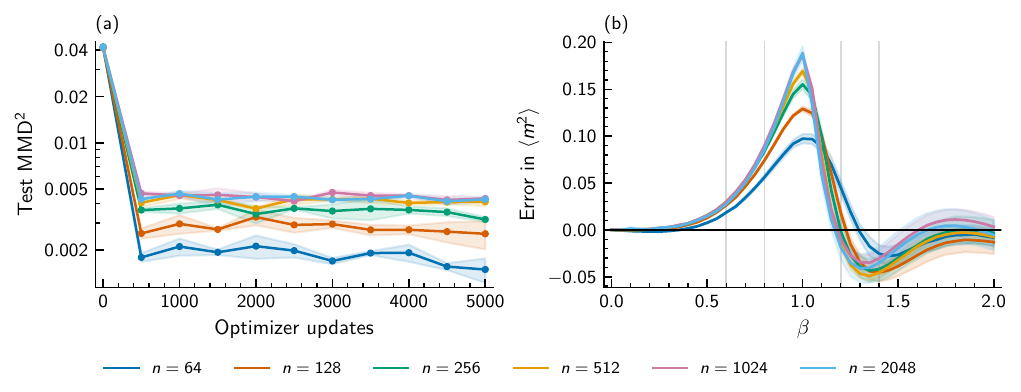}
    \caption{Classical training of conditional IQP Born machines with 64 to 2048 qubits. All models use the parity encoder with the saturating angle of Eq.~\eqref{eq:scaling_schedule} and an open rectangular lattice of trainable one- and two-body phases. (a) Classically estimated test MMD$^2$ during training, averaged over the test temperatures $\beta\in\{0.4,0.95,1.0,1.05,1.6\}$ and two kernel bandwidths whose average bodyness, two and four, is kept fixed as the size varies; logarithmic vertical axis. (b) Error in the classically estimated mean squared magnetization, the model estimate at the selected checkpoint of each fit minus the exact target, at $41$ equally spaced temperatures in $[0,2]$. Gray vertical lines mark the training temperatures. Lines and bands show the mean and one standard deviation across three fits per size.}
    \label{fig:scaling_observables}
\end{figure*}

At every size, the classically estimated test MMD$^2$ decreases within the first $500$ updates and then stays nearly constant, from $0.042$ to $0.0015$ at 64 qubits and to $0.0043$ at 2048 qubits (Fig.~\ref{fig:scaling_observables}(a)). Because the kernels emphasize low-order moments \cite{rudolphTrainabilityBarriersOpportunities2024} and their bandwidths are rescaled with $n$, these values are not comparable across sizes, so we validate them with the mean squared magnetization $\langle m^2\rangle$. It is a two-body quantity, $\langle m^2\rangle=1/n+(1-1/n)\,\overline{C}$, where $\overline{C}$ is the mean of $\langle Z_iZ_j\rangle$ over distinct pairs, and the estimators of Sec.~\ref{sec:ratio-estimator} give it classically at any size (Appendix~\ref{sec:appendix_scaling}).

Figure~\ref{fig:scaling_observables}(b) shows that the models reproduce $\langle m^2\rangle$ away from the crossover at every size, within $0.016$ of the target for $\beta\le0.5$ and within $0.039$ for $\beta\ge1.5$, beyond the training temperatures. The long-range order of the cold phase is therefore learned and extrapolated up to $2048$ qubits. Near the crossover, the learned transition is broader than that of the target, whose finite-size width shrinks as $n$ grows. At $\beta=1$, the model gives $\langle m^2\rangle$ between $0.21$ and $0.24$ at every size, whereas the target decreases from $0.142$ at $64$ qubits to $0.026$ at $2048$ qubits, and the error grows from $0.097$ to $0.189$. The crossover error also appears at the training temperature $\beta=0.8$, where the model exceeds the target by $0.056$ to $0.090$. A 64-qubit model with all-to-all two-body connectivity, selected by a separate search with the same estimator budget, reduces the error at $\beta=1$ from $0.097$ to $0.013$ but underestimates $\langle m^2\rangle$ beyond the development temperatures (Appendix~\ref{sec:appendix_scaling}). Although it also selected different encoder settings, this suggests that the lattice restriction contributes substantially to the crossover error.

We also chose the lattice ansatz because it maps directly onto superconducting processors with square-lattice connectivity. Sampling $16$- and $30$-qubit models on IQM's Emerald quantum computer reproduces the unimodal-to-bimodal transition qualitatively, with substantial errors in the ordered regime, where device noise contracts the ordered peaks and the profile TVD at $\beta=2$ reaches $0.55$ and $0.76$ (Appendix~\ref{sec:hardware}).

\section{Discussion}
\label{sec:discussion}
Proposition~\ref{prop:separation} shows that, at a fixed number of qubits, input conditioning can represent conditional families that no diagonally conditioned IQP model reproduces, because the IQP unitary cannot change the output's $X$-basis distribution. On the two target families studied here, models conditioned through the input learned the family more accurately than diagonal conditioning under matched protocols, even when the diagonal baseline had more parameters. These targets need not lie outside the reach of diagonal conditioning; rather, the experiments indicate that carrying the condition in the input also eases learning in practice. Together, these results extend classically trained IQP Born machines from single distributions to conditional families, with a training cost set by the sampling-and-ratio interface of the encoder.

Within input conditioning, the encoder strongly affects performance. On the Curie--Weiss model, two encoders that impose the same symmetry differ by more than a factor of two in mean TVD, and removing the twist from the selected configuration makes it less accurate than the simpler parity encoder. The encoder also constrains the model before training. The IQP unitary leaves the distribution of outcomes in the $X$ basis unchanged, so at every condition this distribution equals the reference distribution $|\psi_c(z)|^2$ (Eq.~\eqref{eq:x-basis-conservation}). The output state must therefore have the target as its computational-basis distribution and $|\psi_c|^2$ as its $X$-basis distribution. When no $n$-qubit state has this pair of distributions, as for the family of Proposition~\ref{prop:separation} with any fixed input, no choice of phases reproduces the target at that condition. This compatibility is necessary but not sufficient, since one shared IQP unitary must still produce the right phases at every condition. How to test compatibility efficiently, and how to construct tractable encoders that satisfy it for a given target family, remain open questions.

At sixteen qubits, we select models on exact TVD at development conditions, so the reported accuracies characterize the tested configurations rather than a sample-based selection procedure. In a separate sixteen-qubit study, the development score built from the classically estimated MMD had a Spearman rank correlation of $0.97$ with the exact TVD across the checkpoints of $120$ configurations, against $0.83$ for the MMD at the training temperatures (Appendix~\ref{sec:appendix_scaling}), which indicates that sample-based selection can track the exact TVD closely. At large sizes, we validate models through the low-order observable $\langle m^2\rangle$, which, like the low-order emphasis of the MMD kernel, does not certify the full distribution. Both target families are permutation invariant, which makes exact observables tractable at any size but also makes the targets easy to sample classically, so the experiments test representation and generalization rather than a quantum advantage; targets without this symmetry are a natural next step.

Quantum deployment raises separate questions. The sampling-and-ratio interface guarantees efficient classical access to the reference state, not an efficient circuit for preparing $\ket{\phi(c)}$ on hardware. The twist of the twisted parity encoder, like the Curie--Weiss interaction itself, couples every pair of leaves through $(n-1)(n-2)/2$ two-qubit rotations, $105$ at sixteen qubits, which compiles to many gates on processors without all-to-all connectivity. The large-scale and hardware experiments therefore use the parity encoder, which needs at most $2(n-1)$ CNOT gates along a tree of nearest-neighbor couplers. The most accurate sixteen-qubit models and the large-scale demonstrations thus rely on different encoders, which reflects a trade-off between accuracy and preparation cost.

Classical trainability also raises the question of whether the trained models remain hard to sample. When the encoder can be absorbed into the diagonal layer, conditioning reparameterizes a standard IQP circuit, and sampling-hardness results for that circuit family carry over under the same assumptions \cite{bremnerClassicalSimulationCommuting2011,bremnerAverageCaseComplexityApproximate2016}, although they do not certify that a particular trained circuit is hard to sample. Other encoders, including diagonal-equivalent ones with non-uniform reference magnitudes, need not admit this reduction, and the separation shown here concerns expressivity, not sampling complexity. Moreover, efficient estimation of the objective does not guarantee successful optimization: generative models whose output distributions anticoncentrate, a property that hardness arguments typically use, suffer from loss concentration and are not trainable on average \cite{herbstLimitsQuantumGenerative2025}. Characterizing which encoders preserve sampling hardness, and whether training reaches hard instances, remains open.

\section{Conclusions and outlook}
\label{sec:conclusions}

Conditional IQP Born machines encode the condition in the input state and share one IQP unitary across conditions, and they can be trained classically whenever the input state supports efficient sampling and amplitude ratios. Encoding the condition in the input lets the model represent conditional families that no diagonally conditioned IQP model with the same number of qubits reproduces (Proposition~\ref{prop:separation}). In our sixteen-qubit experiments, input conditioning was also more accurate than diagonal conditioning under the same training protocol, with a six times lower mean TVD on the Curie--Weiss model.

We trained conditional IQP Born machines for the Curie--Weiss model with up to $2048$ qubits. The classical estimator used in training also enabled us to estimate physical observables of the trained circuits, such as the mean squared magnetization, at every size without quantum hardware. Because the Curie--Weiss model is permutation invariant, its exact observables reduce to sums over the $n+1$ values of the magnetization and remain computable at these sizes, so we could check what the models learned against the target. Both sides of this comparison are classically accessible, which makes the Curie--Weiss model a case where the workflow can be verified today, at a scale beyond state-vector simulation. The most interesting cases are those that cannot be checked classically, such as physical systems characterized only through experimental measurements. For such systems, a conditional IQP Born machine could still be trained classically at scale on the measured settings, such as temperatures, and then sampled to generate new measurements at other settings, including ones never probed in the experiment. This last step would need a fault-tolerant quantum computer of this size.

Recio-Armengol \textit{et al.} anticipated extending classical training to broader input states \cite[Secs.~6 and 9.5]{recio-armengolTrainClassicalDeploy2026}; the sampling-and-ratio interface realizes this extension and lets the input depend on the condition. Three directions follow. The first is to characterize which conditional families a class of tractable encoders can represent, how this compares with adding hidden qubits \cite{kurkinUniversalityKernelAdaptive2025}, and how the encoder shapes generalization, as it does for supervised quantum models \cite{caroEncodingdependentGeneralizationBounds2021}. The second is to learn the encoder jointly with the IQP unitary, which requires extending the classical estimators to derivatives with respect to the input state; such trainable inputs would also benefit unconditional IQP Born machines. The third is to use conditional IQP Born machines as classically trainable components of generative architectures that reduce to sequences of condition-dependent sampling tasks, such as quantum denoising diffusion models \cite{cacioppoQuantumDiffusionModels2023,quinnConditioningGenerativeQuantum2026,chenOvercomingDimensionalFactorization2026} and autoregressive models.

\section*{Data availability}
The code and data supporting this study are available on GitHub \cite{balloConditionalGenerationIQP2026}.

\section*{Author contributions}
O.B. conceived the study, developed the formulation, carried out the experiments, and wrote the manuscript. The idea emerged from discussions between O.B. and E.T., whose subsequent input shaped the study and the manuscript. L.D. critically reviewed the manuscript, in particular its theoretical results. F.V. and A.G. supervised the research throughout the project, providing scientific guidance and oversight. All authors reviewed the final manuscript.

Large language models were used to assist with implementing the experiment and plotting code and with checking the grammar of the manuscript. All scientific content was conceived, verified, and approved by the authors, who take full responsibility for it.

\section*{Acknowledgments}
O.B. is a fellow of Eurecat’s ``Vicente L\'opez'' PhD grant program. 
This work was partially supported by grants PID2024-157778OB-I00 and AIA2025-163435-C43 from the Spanish Ministry of Science and Innovation, the European Union, and the Agencia Estatal de Investigaci\'on, and grant No.~TSI-100929-2023-2, C\'atedra ENIA UAB-Cru\"{\i}lla, from the Spanish Ministry of Economic Affairs and Digital Transformation and the European Union NextGenerationEU.

\bibliography{references}

\clearpage
\onecolumngrid
\appendix
\newcommand{\hpCell}[2]{\parbox[t]{#1}{\raggedright #2\strut}}
\raggedbottom

\section{Proof of Proposition~\ref{prop:separation} and estimator variance}
\label{sec:appendix_theory}

This appendix contains the proof of Proposition~\ref{prop:separation} (Sec.~\ref{sec:gauge}) and a bound on the variance of the training estimator, which supports the estimation claims of Secs.~\ref{sec:model-access} and \ref{sec:unbiased-conditional-mmd}.

\subsection{Proof of Proposition~\ref{prop:separation}}
\label{sec:appendix_two_qubit_example}

\emph{Part (i).} By Eq.~\eqref{eq:x-basis-conservation}, every output of diagonal conditioning on a fixed input $\ket{\phi}$ has $X$-basis distribution $|\psi(z)|^2$, the reference distribution of $\ket{\psi}=H^{\otimes n}\ket{\phi}$, whatever the parameters and generators. A member $p(\cdot\mid c)$ is therefore reproduced only if some pure state has computational-basis distribution $p(\cdot\mid c)$ and $X$-basis distribution $|\psi(z)|^2$, that is, only if the two are compatible. Conversely, a diagonal layer with all $2^n$ Pauli-$Z$ strings can multiply each entry of $\ket{\psi}$ by any phase and hence reach every such state, so the condition is then also sufficient.\qed

\emph{Part (ii).} Fix $n\geq2$ and an IQP unitary $U_{\boldsymbol\theta}$, and write $e_j$ for the string with a single one at position $j$. Consider a family of two distributions, indexed by $c\in\{0,1\}$: the first always outputs $0^n$, and the second outputs $0^n$, $e_1$, or $e_2$ with probability $\tfrac13$ each. We show that input conditioning reproduces this family with any fixed IQP unitary, whereas no fixed input does. The two distributions are the computational-basis distributions of
\begin{equation}
    \ket{\omega_0}=\ket{0^n},\qquad
    \ket{\omega_1}=\tfrac1{\sqrt3}\bigl(\ket{0^n}+\ket{e_1}+\ket{e_2}\bigr).
    \label{eq:two-member-states}
\end{equation}

\emph{Input conditioning reproduces the family.} Choose the encoder so that the reference state at condition $c$ is
\begin{equation}
    \ket{\psi_c}=D(\boldsymbol\theta)^\dagger H^{\otimes n}\ket{\omega_c}.
\end{equation}
The final Hadamard layer in the IQP unitary gives the output $U_{\boldsymbol\theta}\ket{\phi(c)}=H^{\otimes n}D(\boldsymbol\theta)\ket{\psi_c}=\ket{\omega_c}$. Since $D(\boldsymbol\theta)^\dagger$ only multiplies each amplitude by a phase, $|\psi_c(z)|=|(H^{\otimes n}\omega_c)(z)|$, where $(H^{\otimes n}\omega_c)(z)=\bra zH^{\otimes n}\ket{\omega_c}$. Because $\ket{\omega_c}$ is a two-qubit state tensored with $\ket0^{\otimes(n-2)}$, $H^{\otimes n}\ket{\omega_c}$ is a two-qubit state tensored with $\ket{+}^{\otimes(n-2)}$. These reference states therefore admit the sampling-and-ratio interface. Sampling the reference distribution amounts to drawing the first two bits from this two-qubit state and the remaining $n-2$ bits uniformly, and the amplitude ratios factorize as
\begin{equation}
    R_c(z,a)=e^{-i\Delta_af_{\boldsymbol\theta}(z)}\,
    \frac{(H^{\otimes n}\omega_c)(z\oplus a)}{(H^{\otimes n}\omega_c)(z)},
\end{equation}
where the fraction depends only on the first two bits of $z$ and $a$, and the phase factor is the one that the estimator already evaluates, in $O(|\mathcal G|)$ operations, Eq.~\eqref{eq:moment-as-reference-average}.

\emph{No fixed input reproduces it.} By part (i), a fixed input would need an $X$-basis distribution compatible with both members. The only state with the first distribution is $\ket{0^n}$, up to a global phase, and its $X$-basis distribution is uniform (Sec.~\ref{sec:gauge}). It therefore suffices to show that no state with the second distribution has a uniform $X$-basis distribution. Such a state has the form
\begin{equation}
    \ket\omega=A\ket{0^n}+B\ket{e_1}+C\ket{e_2},
    \qquad |A|=|B|=|C|=\tfrac1{\sqrt3}.
\end{equation}
If its $X$-basis distribution $P_X$ were uniform, every expectation value $\langle X_a\rangle=\sum_zP_X(z)(-1)^{a\cdot z}$ with $a\neq0$ would vanish, because $(-1)^{a\cdot z}$ equals $+1$ and $-1$ for equally many strings $z$. Since $X_1$ flips the first bit, $X_1\ket\omega=B\ket{0^n}+A\ket{e_1}+C\ket{e_1\oplus e_2}$, and as $\ket\omega$ has no component along $\ket{e_1\oplus e_2}$, $\langle X_1\rangle=\bar AB+\bar BA$. The same computation for $X_2$ and $X_1X_2$ gives
\begin{equation}
    \langle X_1\rangle=2\operatorname{Re}(\bar AB),\qquad
    \langle X_2\rangle=2\operatorname{Re}(\bar AC),\qquad
    \langle X_1X_2\rangle=2\operatorname{Re}(\bar BC).
    \label{eq:two-member-moments}
\end{equation}
The global phase of $\ket\omega$ does not affect these values, so we may take $A$ real and positive. The first two then vanish only if $B$ and $C$ are purely imaginary, which makes $\bar BC$ real and nonzero, so the third cannot vanish. No $X$-basis distribution is compatible with both members, and no fixed input reproduces the family. The same argument applies to any distribution supported on three strings; for $n=2$, the second member is the distribution of Ref.~\cite[Lemma~11]{kurkinUniversalityKernelAdaptive2025}.\qed

\subsection{Variance of the estimator}
\label{sec:estimator-variance}

Section~\ref{sec:model-access} states that the objective and its gradients can be estimated to additive accuracy $\epsilon$ in time polynomial in $n$ and $1/\epsilon$, and Sec.~\ref{sec:unbiased-conditional-mmd} that the variances of the estimators are bounded independently of $n$. Recio-Armengol \textit{et al.} bound the variance of a single sampled expectation value by the inverse number of samples \cite[Sec.~4.1]{recio-armengolTrainClassicalDeploy2026}, and Sec.~\ref{sec:ratio-estimator} gives the same bounds for the ratio estimators. The complete estimator of Eq.~\eqref{eq:sampled-mmd-estimator} combines such terms on shared samples, and the same argument bounds it directly.

For each condition and mask, write $\widehat{\delta_c^2}(a)=U_Y-2\overline Y\,\overline T+U_T$ for the three terms of Eq.~\eqref{eq:conditional-mmd-ustatistic}. Since $|T_b|=1$, we have $|U_T|\leq1$ and $\|\overline Y\,\overline T\|_{L^2}\leq\|\overline Y\|_{L^2}\leq1$. Since distinct samples are independent, $\|Y_sY_t\|_{L^2}=\|Y_s\|_{L^2}\|Y_t\|_{L^2}\leq1$ and $\|Y_sY'_t\|_{L^2}\leq2$ for $s\neq t$, and averaging preserves these bounds, so $\|U_Y\|_{L^2}\leq1$ and the mixed pair mean of Eq.~\eqref{eq:conditional-mmd-gradient} is bounded by $2$. The triangle inequality then gives
\[
\begin{aligned}
    \E\!\left[\bigl(\widehat{\delta_c^2}(a)\bigr)^2\right]&\leq16,\\
    \E\!\left[\bigl(\partial_{\theta_g}\widehat{\delta_c^2}(a)\bigr)^2\right]&\leq64,
\end{aligned}
\]
for every $n$. Since $0<Q_\sigma^+\leq1$, averaging over conditions, bandwidths, and masks preserves these bounds. Averaging $N$ independent repetitions of the complete estimator therefore gives mean-square error at most $16/N$ for the objective and $64/N$ for each gradient component, and by Chebyshev's inequality, $O(\epsilon^{-2})$ repetitions give additive error $\epsilon$ at fixed confidence.

\section{Training and experiment details}
\label{sec:appendix_protocols}

This appendix gives the settings behind the results of Sec.~\ref{sec:results}. Appendix~\ref{sec:appendix_hyperparams} lists the settings common to the IQP experiments, and each following subsection gives the details of one experiment, titled after the corresponding section, followed by the deployment on quantum hardware. The accompanying repository contains the code, the frozen protocol file of every experiment, with the complete search spaces, condition grids, estimator budgets, and random streams, and the result bundles behind every figure and table \cite{balloConditionalGenerationIQP2026}. Running \texttt{reproduce.py} retrains every reported model from the selected configurations and regenerates all figures and tables; with \texttt{-{}-full}, it also reruns the hyperparameter searches. The hardware samples are read from the stored data, as device noise and calibration drift prevent exact reproduction.

\subsection{Common training settings}
\label{sec:appendix_hyperparams}

All IQP models are trained with \texttt{Adam} \cite{kingmaAdam2015}, with $\epsilon=10^{-8}$ and no weight decay. Each update draws one training condition and one kernel bandwidth uniformly. For an average bodyness $\bar\ell$, we set the bandwidth by inverting $\bar\ell=np_\sigma$ with Eq.~\eqref{eq:mmd-mask-law}, which gives
\begin{equation}
    \sigma(n,\bar\ell)=\left[-2\ln\left(1-\frac{2\bar\ell}{n}\right)\right]^{-1/2},
    \qquad 0<\bar\ell<n/2,
\end{equation}
where $\bar\ell$ counts the identity mask, as in Sec.~\ref{sec:unbiased-conditional-mmd}. Table~\ref{tab:training_settings} lists the training settings and search spaces of each experiment, and Table~\ref{tab:selected_hyperparameters} the selected hyperparameters. At sixteen qubits, we select checkpoints and hyperparameters on the exact TVD at development conditions outside the training set; the per-temperature circuits of Sec.~\ref{sec:cw_main} are selected at the temperatures at which they are trained. The large-scale and hardware models are selected on the classically estimated MMD (Appendix~\ref{sec:appendix_scaling}).

\begin{table*}[t]
\centering
\caption{Training settings and hyperparameter search spaces. Searched hyperparameters list their range; ``log'' marks a log-uniform prior. Cosine schedules decay to $0.1$ times the initial learning rate unless the final ratio is searched.}
\label{tab:training_settings}
\footnotesize
\setlength{\tabcolsep}{6.5pt}
\begin{tabular}{@{}llll@{}}
\toprule
 & Hamming-weight family & Curie--Weiss, $n=16$ & Large-scale training \\
\midrule
Updates (checkpoint interval) & $2000$ ($100$) & $2000$ ($100$) & $5000$ ($250$) \\
Training conditions & nine, on the two axes & $\beta\in\{0.1,0.7,1.3,1.95\}$ & $\beta\in\{0.6,0.8,1.2,1.4\}$ \\
Target pool; minibatch & $1024$; $256$ & $1024$; $256$ & $10\,000$; $450$ \\
Masks; reference samples & $1024$; $1024$ & $2048$; $1024$ & searched \\
Average bodyness $\bar\ell$ & $2$, $\bar\ell_2$, $\bar\ell_3$ & $2$, $\bar\ell_2$, $\bar\ell_3$ & $2$, $4$ \\
\midrule
Search & $60$ random $+$ $60$ TPE trials & $60$ random $+$ $60$ TPE trials & $60$ random trials per schedule \\
Learning rate & $[3\times10^{-4},0.1]$, log & $[3\times10^{-4},0.1]$, log & $[10^{-4},0.05]$, log \\
Schedule & constant or cosine & constant or cosine & cosine, final ratio $[0.05,1]$, log \\
$\beta_1$ & $[0.8,0.95]$ & $[0.8,0.95]$ & $[0.8,0.95]$ \\
$1-\beta_2$ & $[10^{-3},0.05]$, log & $[10^{-3},0.05]$, log & $[10^{-3},0.05]$ \\
Initialization width & $[0.01,1]$, log & $[0.01,1]$, log & $[0.005,0.3]$, log \\
One-body mean $\bar\theta$ & $[0,\pi/2]$ & $0$ & $0$ \\
Kernel & $\bar\ell_2\leq\bar\ell_3\in\{4,4.5,\dots,7.5,7.99\}$ & $\bar\ell_2\leq\bar\ell_3\in\{4,4.5,\dots,7.5,7.99\}$ & --- \\
Masks; references at $n=64$ & --- & --- & $\{512,1024,2048\}$; $\{1024,2048\}$ \\
Budget growth & --- & --- & $(n/64)^{\alpha}$, $\alpha\in\{0,\tfrac14,\tfrac12\}$ \\
Encoder constants & --- & $\lambda\in[0.25,1.5]$ or $\eta\in[0.3,0.7]$, log & \hpCell{0.24\textwidth}{$t_{\max}\in[0.5,\pi]$, $\tau\in[1,20]$ log, $\beta_0\in[0.5,1.5]$, $\gamma\in[0,0.5]$} \\
\bottomrule
\end{tabular}
\end{table*}

\begin{table*}[t]
\centering
\caption{Selected hyperparameters, rounded to two significant figures; the protocol files in the repository give the exact values. The large-scale configuration is used at every size and for the hardware models. For the classical models, the initialization width is the multiplier of the Xavier initialization.}
\label{tab:selected_hyperparameters}
\footnotesize
\setlength{\tabcolsep}{5pt}
\begin{tabular}{@{}llllllll@{}}
\toprule
Model & Learning rate & Schedule & Init.\ width & $\beta_1$ & $1-\beta_2$ & $\bar\ell_2$, $\bar\ell_3$ & Other \\
\midrule
Hamming-weight, input conditioning & $0.0012$ & cosine & $0.011$ & $0.83$ & $0.0011$ & $5$, $7$ & $\bar\theta=1.2$ \\
Hamming-weight, diagonal conditioning & $0.00070$ & cosine & $0.019$ & $0.89$ & $0.014$ & $4.5$, $7.99$ & $\bar\theta=0.83$ \\
\addlinespace
C-IQP (TP) & $0.0055$ & cosine & $0.076$ & $0.94$ & $0.0013$ & $5$, $7$ & $\eta=0.47$ \\
C-IQP (P) & $0.0034$ & cosine & $0.022$ & $0.80$ & $0.0080$ & $4$, $7.5$ & $\lambda=0.68$ \\
D-IQP & $0.0091$ & constant & $0.10$ & $0.85$ & $0.0027$ & $4$, $5$ & --- \\
Per-$\beta$ IQP (TP) & $0.0051$ & constant & $0.073$ & $0.82$ & $0.028$ & $5$, $7$ & $\eta=0.45$ \\
Per-$\beta$ IQP & $0.090$ & cosine & $0.033$ & $0.95$ & $0.0012$ & $5$, $5.5$ & --- \\
\addlinespace
Small RBM & $0.073$ & cosine & $1.2$ & $0.82$ & $0.048$ & --- & --- \\
Small Transformer & $0.040$ & cosine & $0.39$ & $0.95$ & $0.010$ & --- & --- \\
Small VAE & $0.041$ & cosine & $0.13$ & $0.83$ & $0.0022$ & --- & --- \\
Large RBM & $0.045$ & constant & $0.12$ & $0.86$ & $0.0026$ & --- & --- \\
Large Transformer & $0.0031$ & cosine & $0.22$ & $0.94$ & $0.036$ & --- & --- \\
Large VAE & $0.039$ & cosine & $1.8$ & $0.88$ & $0.0053$ & --- & --- \\
\addlinespace
Large-scale, all-to-all ($n=64$) & $0.0017$ & cosine & $0.013$ & $0.93$ & $0.018$ & --- & \hpCell{0.17\textwidth}{$t_{\max}=1.2$, $\tau=4.1$, $\beta_0=1.0$, $\gamma=0.19$} \\
Large-scale, lattice & $0.041$ & cosine & $0.082$ & $0.90$ & $0.043$ & --- & \hpCell{0.17\textwidth}{$t_{\max}=1.6$, $\tau=2.6$, $\beta_0=1.2$, $\gamma=0.04$} \\
\bottomrule
\end{tabular}
\par\smallskip
\begin{minipage}{0.92\textwidth}\footnotesize
The large-scale configuration uses a final learning-rate ratio of $0.09$, $512$ masks and $1024$ reference samples at $n=64$, and budget growth $\alpha=\tfrac12$. The all-to-all configuration uses a final learning-rate ratio of $0.34$ and the same $512$ masks and $1024$ reference samples.
\end{minipage}
\end{table*}

\subsection{Details for generalization to unseen condition combinations}
\label{sec:gaussian-two-dimensional-protocol}

\emph{Conditions.} Training uses the nine conditions $(u,0)$ at $u\in\{-1,\allowbreak -0.37,\allowbreak 0,\allowbreak 0.37,\allowbreak 1\}$ and $(0,v)$ at $v\in\{-1,\allowbreak -0.31,\allowbreak 0.31,\allowbreak 1\}$. Development uses $(u,0)$ at $u\in\{\pm0.17,\pm0.73\}$ and $(0,v)$ at $v\in\{\pm0.13,\pm0.69\}$. The sixteen mixed conditions $\{-1,-0.37,0.37,1\}\times\{-1,-0.31,0.31,1\}$, with both coordinates nonzero, are used for reporting; four lie inside the training convex hull $|u|+|v|\leq1$, and twelve lie outside.

\emph{Training and selection.} All models use the settings of Table~\ref{tab:training_settings} and are selected by the mean exact TVD over the development conditions. Table~\ref{tab:selected_hyperparameters} gives the selected configurations.

\emph{Encoder evaluation.} The encoded input and its Hadamard transform are invariant under qubit permutations, so the input and reference amplitudes depend only on Hamming weight. The reference amplitudes follow from the input amplitudes by an $(n+1)\times(n+1)$ linear transform over Hamming weights (a Krawtchouk transform), computed once per condition in $O(n^3)$ operations. A reference bitstring is sampled by drawing its weight and then a uniformly random string of that weight, and each amplitude ratio is a quotient of two table entries, costing $O(n)$.

\emph{Diagonal conditioning.} The baseline applies the same lattice circuit to $\ket{0}^{\otimes16}$, with phases $\boldsymbol\theta(u,v)=\mathbf w_0+u\,\mathbf w_u+v\,\mathbf w_v$. We initialize $\mathbf w_u=\mathbf w_v=0$ and draw $\mathbf w_0$ as the phases of the conditional model. Three variants use the quadratic map $\boldsymbol\theta(u,v)=\mathbf w_0+u\,\mathbf w_u+v\,\mathbf w_v+u^2\mathbf w_{uu}+v^2\mathbf w_{vv}$ on the lattice ($200$ parameters), and the affine and quadratic maps on all $136$ one- and two-body phases ($408$ and $680$ parameters). All four follow the same protocol.

\emph{Results.} Table~\ref{tab:gaussian-two-dimensional} lists the TVDs at the mixed conditions. At the origin, the affine diagonal baseline has TVD $0.017\pm0.002$. Averaged over the ten fits, it reproduces between $0.95$ and $1.02$ of the target change in mean along $u$, for $v\in\{-1,0,1\}$, and between $0.03$ and $0.04$ of the target change in width along $v$, for $u\in\{-1,0,1\}$. Across the $41\times41$ grid of Fig.~\ref{fig:gaussian-two-dimensional}(a), the TVD of the conditional model has correlation $0.91$ with the TVD between each target and the uniform distribution, and $0.72$ with the distance from each condition to the nearest training condition.

\begin{table}[!t]
\caption{Two-coordinate errors after axis training, evaluated on the sixteen mixed conditions. Input conditioning is the conditional IQP Born machine; diagonal conditioning makes the phases of the same lattice circuit, acting on $\ket{0}^{\otimes16}$, an affine function of the condition, and its variants use a quadratic condition map, an all-to-all diagonal layer, or both. Values are means and standard deviations of the per-seed mean full-distribution TVD across ten seeds; the deterministic encoder-only row has no spread. Lower is better.}
\label{tab:gaussian-two-dimensional}
\centering
\begin{tabular}{lc}\toprule
Model & TVD\\\midrule
Encoder only & $0.387$\\
Input conditioning (40 parameters) & $0.0216\pm0.0017$\\
Diagonal conditioning (120 parameters) & $0.069\pm0.002$\\
Diagonal, quadratic map (200 parameters) & $0.078\pm0.002$\\
Diagonal, all-to-all layer (408 parameters) & $0.103\pm0.004$\\
Diagonal, all-to-all layer, quadratic map (680 parameters) & $0.109\pm0.005$\\
\bottomrule\end{tabular}
\end{table}

\subsection{Details for conditional modeling of the Curie--Weiss model}
\label{sec:appendix_cw}

All models use the same IQP ansatz, with sixteen one-body and $120$ two-body phases and no ancillary qubits, and the encoders of Eqs.~\eqref{eq:cw_twisted_input} and \eqref{eq:cw_parity_input}, or $\ket{0}^{\otimes16}$ for per-$\beta$ IQP.

\subsubsection{Twisted parity encoder}
\label{sec:appendix_twisted}

Every layer of the twisted parity encoder is invariant under permutations of the fifteen leaves, so the input and reference amplitudes depend only on the root bit and the leaf Hamming weight. The leaf state stays in the $16$-dimensional symmetric subspace spanned by Dicke states, where the twist is diagonal and the collective rotation follows from the collective spin operators, so a $16\times16$ matrix exponential gives the twisted leaf state. The Hamming-weight transform of Appendix~\ref{sec:gaussian-two-dimensional-protocol} then gives the $2\times16$ table of reference amplitudes in $O(n^3)$ operations per temperature. Sampling draws the root bit and leaf weight from this table and then a uniformly random set of leaves; since a mask changes the leaf weight to $w+|a_L|-2\,z_L\cdot a_L$ and the root bit to $b\oplus a_{16}$, where $z_L$ and $a_L$ are the leaf parts of the sampled string and the mask, each amplitude ratio is a quotient of two table entries and costs $O(n)$.

To prepare the encoder, we apply a Hadamard gate to every qubit, an $R_{ZZ}(2\mu_\beta/15)$ rotation to every pair of leaves, an $R_X(-\pi/4)$ rotation to every leaf, and an $R_{ZZ}(t_\beta)$ rotation between each leaf and the root. The $105$ leaf-pair rotations commute, so the preparation is exact and needs no Trotterization.

We fix $\kappa=0.4$ and the rotation angle $\pi/4$, and $\eta$ is searched in $[0.3,0.7]$ (Appendix~\ref{sec:cw_hpo}). For one-axis twisting, the rotation angle that narrows the leaf Hamming-weight distribution most tends to $\pi/4$ as the twisting strength vanishes \cite{kitagawaSqueezedSpinStates1993}; for the selected schedule, it lies between $36^\circ$ and $40^\circ$ over $\beta\in[0.5,2]$.

\subsubsection{Parity encoder}

For a reference bitstring $(r,b)$, with leaf bits $r\in\{0,1\}^{15}$ and root bit $b\in\{0,1\}$, the parity encoder in Eq.~\eqref{eq:cw_parity_input} with angle $t$ gives the amplitude
\begin{equation}
 \psi_{\mathrm P}(r,b)=
 \mathbf{1}_{b=\oplus_i r_i}
 \cos(t/2)^{15-|r|}
 [-i\sin(t/2)]^{|r|}.
\end{equation}
C-IQP (P) uses $t=\lambda\beta$. To sample this state, we draw the first fifteen bits independently with probability $\sin^2(t/2)$ of obtaining one, and set the root bit to their parity. For an even-parity mask, the amplitude ratio is the product of the corresponding single-bit amplitude ratios over the leaves; for an odd-parity mask, it vanishes. Both operations require $O(n)$ arithmetic operations. We prepare the reference state by applying $R_X(t)$ to each leaf and a CNOT from each leaf to the root. A final Hadamard layer prepares the encoded input.

\subsubsection{Training and selection}
\label{sec:cw_hpo}

We fit the shared models at $\beta\in\{0.1,0.7,1.3,1.95\}$. Each per-temperature family contains three circuits fitted directly at $\beta\in\{0.4,1,2\}$, with separate learned phases and one jointly selected configuration.

We select checkpoints and configurations using a full-distribution TVD score $E_{\mathrm{TVD}}$ and a physical-observable score $E_{\mathrm{phys}}$. For a temperature grid $\Gamma$ and a nonnegative error curve $e$, define
\begin{equation}
 A_\Gamma(e)=\frac{2\max_{\beta\in \Gamma}e(\beta)+\overline e_\Gamma}{3},
\end{equation}
where $\overline e_\Gamma$ is the trapezoidal integral over the grid divided by its temperature span. Averages over the evaluation grid use this normalized trapezoidal rule, and averages over the three evaluation temperatures are arithmetic means. We compute $E_{\mathrm{TVD}}=A_\Gamma(\mathrm{TVD})$ from exact probabilities over all $2^{16}$ bitstrings, and
\begin{equation}
 E_{\mathrm{phys}}=\frac{A_\Gamma(e_m)+A_\Gamma(e_c)}{2},\qquad
 e_m=\left|\langle|m|\rangle_q-\langle|m|\rangle_p\right|,
 \label{eq:cw_selection_errors}
\end{equation}
with $e_c=|c_{\mathrm{fluc},q}-c_{\mathrm{fluc},p}|/c_\star$, where $c_\star\approx0.823$ is the maximum target value of $c_{\mathrm{fluc}}$ for $0\leq\beta\leq2$ (Appendix~\ref{sec:cw_physical_observables}). Shared models are scored on a grid $\Gamma_{\mathrm{sel}}$ of seventeen temperatures in $[0.05,1.975]$ that excludes the evaluation temperatures $\beta\in\{0.4,1,2\}$ but contains a point within $0.025$ of each, and each per-temperature circuit is scored at the temperature at which it is trained. Within each fit, we select the checkpoint with the smallest $E_{\mathrm{phys}}$ among those whose $E_{\mathrm{TVD}}$ lies within $0.01$ of the minimum.

The TPE sampler receives only $E_{\mathrm{TVD}}$. Confirmation refits the top three or four configurations with five fits of three initializations each, ranking shared models on $\Gamma_{\mathrm{sel}}$ together with twenty interleaved temperatures, and the final stage performs ten fits with three initializations per circuit. All families share the random streams for target pools and stochastic estimates. The evaluation grid contains $101$ temperatures in $[0,2]$; neither this grid nor $\beta\in\{0.4,1,2\}$ enters the training or selection of the shared models.

\subsubsection{Diagonal conditioning}
\label{sec:cw_diagonal}

The D-IQP baseline uses the fixed input $\exp[-i(\pi/4)\sum_{i=1}^{15}Z_iZ_{16}]\ket{+}^{\otimes16}$, the parity encoder at $t=\pi/2$, whose reference state is uniform on even-parity strings, so the output is invariant under global spin flips for every parameter vector. Its phases $\boldsymbol\theta(\beta)=\sum_{d=0}^{3}T_d(\beta-1)\mathbf w_d$, with $\mathbf w_d\in\mathbb R^{136}$, map $\beta\in[0,2]$ to the Chebyshev interval $[-1,1]$. We initialize $\mathbf w_0$ from the paired initialization streams of the other models and set $\mathbf w_d=0$ for $d\geq1$, so the initial model is a constant-parameter IQP circuit. It follows the protocol of the other shared models. Over the ten final fits, $E_{\mathrm{TVD}}$ is $0.36\pm0.06$, the temperature-averaged TVD is $0.28\pm0.03$, and the mean TVD lies between $0.22$ and $0.34$ at every grid point.

\subsubsection{Physical observables}
\label{sec:cw_physical_observables}

\begin{table}[!htbp]
\centering
\caption{Observable errors of the sixteen-qubit Curie--Weiss models, averaged over $\beta\in\{0.4,1,2\}$ and over ten fits: $e_m$ is the absolute error in $\langle|m|\rangle$ and $e_c$ the absolute error in $c_{\mathrm{fluc}}$ divided by the maximum target value, Eq.~\eqref{eq:cw_selection_errors}.}
\label{tab:cw_observable_errors}
\begin{tabular}{lcc}
\toprule
Model & $e_m$ & $e_c$ \\
\midrule
C-IQP (TP) & $0.007$ & $0.010$ \\
C-IQP (P) & $0.028$ & $0.058$ \\
D-IQP & $0.056$ & $0.32$ \\
Per-$\beta$ IQP (TP) & $0.005$ & $0.018$ \\
Large Transformer & $0.006$ & $0.014$ \\
\bottomrule
\end{tabular}
\end{table}

For a distribution $q$, let $\rho_q(k\mid\beta)$ denote its exact Hamming-weight profile and set $m_k=(2k-n)/n$ with $n=16$. We compute its moments as $\langle m^r\rangle_q=\sum_{k=0}^{n}\rho_q(k\mid\beta)m_k^r$ and its mean absolute magnetization as $\sum_k\rho_q(k\mid\beta)|m_k|$. With $H=-nm^2/2$ in units $J=k_B=1$, Eq.~\eqref{eq:cw_heat_fluctuation} becomes
\begin{equation}
    c_{\mathrm{fluc},q}(\beta)=\frac{\beta^2n}{4}
    \left(\langle m^4\rangle_q-\langle m^2\rangle_q^2\right).
\end{equation}
We compute each observable within each fit before averaging.

\subsubsection{Twist ablation}
\label{sec:cw_product_ablation}

To isolate the effect of the twist, we refit the selected C-IQP (TP) configuration with $\kappa=0$, which removes the twist and leaves every other setting unchanged.

\emph{Protocol.} With $\kappa=0$, the twist strength $\mu_\beta$ vanishes, the rotation acts trivially, and the encoder becomes the parity encoder with the schedule $\sin t_\beta=\tanh(\eta\beta^2)$. We follow the final-stage protocol, with ten fits paired with the reported fits, and keep the optimizer settings selected for $\kappa=0.4$ without a new search. We repeat the refit at the two ends of the search range of $\eta$.

\emph{Results.} At the selected $\eta=0.47$, removing the twist raises the temperature-averaged TVD from $0.046\pm0.004$ to $0.132\pm0.004$, above the $0.103\pm0.004$ of C-IQP (P). Every fit has a higher $E_{\mathrm{TVD}}$ and a higher temperature-averaged TVD than its paired C-IQP (P) fit, and hence than its paired C-IQP (TP) fit. At $\eta=0.3$ and $0.7$, the TVD is $0.299\pm0.006$ and $0.148\pm0.009$. Under the tested configurations, the improvement of C-IQP (TP) therefore comes from the twist, and the schedule alone does not reproduce it.

\subsection{Details for comparison with classical models}
\label{sec:cw_classical_details}

All classical models take the inverse temperature $\beta$ as their condition. We symmetrize each model as $q_{\mathrm{sym}}(x\mid\beta)=[q(x\mid\beta)+q(\bar x\mid\beta)]/2$, where $\bar x$ flips every bit, and use $q_{\mathrm{sym}}$ in training, evaluation, and sampling, so every model satisfies $q(x\mid\beta)=q(\bar x\mid\beta)$; none imposes permutation symmetry.

\begin{table*}[t]
\centering
\small
\caption{Comparison of C-IQP models with classical conditional models on the $101$-point Curie--Weiss evaluation grid. All models share the training temperatures, target samples, update budget, hyperparameter search, and selection rule. C-IQP models train on the sampled MMD objective, whereas the classical models train on their native likelihood or evidence lower bound. Parameters count all learned parameters used in training; for the VAEs, parentheses give the generator parameters alone. Mean TVD is the temperature-averaged full-distribution TVD, and Appendix~\ref{sec:cw_hpo} defines the scores $E_{\mathrm{TVD}}$ and $E_{\mathrm{phys}}$. Values are means $\pm$ standard deviations over ten independent fits. Lower is better, and bold marks the lowest mean in each column.}
\label{tab:cw_classical}
\setlength{\tabcolsep}{9pt}
\begin{tabular}{lrrrr}
\toprule
Model & Parameters & $E_{\mathrm{TVD}}$ & Mean TVD & $E_{\mathrm{phys}}$ \\
\midrule
C-IQP (TP) & 136 & $\mathbf{0.060}\pm0.004$ & $\mathbf{0.046}\pm0.004$ & $0.0237\pm0.0011$ \\
C-IQP (P) & 136 & $0.128\pm0.003$ & $0.103\pm0.004$ & $0.086\pm0.009$ \\
\midrule
Small RBM & 140 & $0.16\pm0.06$ & $0.13\pm0.05$ & $0.10\pm0.07$ \\
Small Transformer & 137 & $0.10\pm0.02$ & $0.068\pm0.015$ & $0.050\pm0.016$ \\
Small VAE & 134 (92) & $0.20\pm0.02$ & $0.138\pm0.012$ & $0.169\pm0.018$ \\
\midrule
Large RBM & 986 & $0.076\pm0.009$ & $0.061\pm0.007$ & $0.030\pm0.011$ \\
Large Transformer & 971 & $0.070\pm0.008$ & $0.052\pm0.005$ & $\mathbf{0.022}\pm0.006$ \\
Large VAE & 950 (548) & $0.14\pm0.04$ & $0.10\pm0.03$ & $0.12\pm0.04$ \\
\bottomrule
\end{tabular}
\end{table*}

The RBM has $h$ hidden units and visible and hidden biases that depend affinely on $\beta$, with $h=6$ for the small model and $h=53$ for the large one, and we compute its partition function exactly over the $2^{16}$ visible configurations. The Transformer is a decoder-only model with one block and one attention head that generates the bits autoregressively, with an affine embedding of $\beta$ and model and feed-forward widths $3$ and $7$ for the small model and $8$ and $34$ for the large one. The VAE has a one-dimensional latent variable with a standard normal prior and single-hidden-layer encoder and decoder networks that receive $\beta$ by concatenation, with widths $2$ and $4$ for the small model and $20$ and $28$ for the large one. We symmetrize its Bernoulli decoder likelihood as the equal mixture of the probabilities of $x$ and $\bar x$ and train on the evidence lower bound of this symmetrized decoder. Parameter counts include the VAE encoder, which is needed only for training; the generators alone have $92$ and $548$ parameters. The RBM and Transformer minimize the exact negative log-likelihood, and we evaluate their exact probabilities. For the VAE, we integrate the decoder over the latent prior by Gauss--Hermite quadrature, refined until the TVD changes by at most $10^{-5}$.

The classical models reuse the training temperatures, target pools, minibatch size, update budget, checkpoint schedule, \texttt{Adam} optimizer, stages, scores, and tolerance of the IQP models, and fits with the same index share their target samples, which allows the paired comparisons of Sec.~\ref{sec:cw_classical}. Each model family has its own search of $60$ random and $60$ TPE trials over the learning rate, schedule, and \texttt{Adam} moments, with the Curie--Weiss ranges of Table~\ref{tab:training_settings}, and over a multiplier in $[0.1,3]$, log-uniform, that scales the Xavier initialization of the weights; biases start at zero. Table~\ref{tab:selected_hyperparameters} gives the selected configurations.

\subsection{Details for large-scale training}
\label{sec:appendix_scaling}

\emph{Encoder and architecture.} We use the parity encoder in Eq.~\eqref{eq:cw_parity_input}, generalized to $n$ qubits, with the saturating angle
\begin{equation}
 t_n(\beta)=t_{\max}\,\frac{\operatorname{sig}\!\left(\tau_n(\beta-\beta_0)\right)-\operatorname{sig}(-\tau_n\beta_0)}{1-\operatorname{sig}(-\tau_n\beta_0)},
 \label{eq:scaling_schedule}
\end{equation}
where $\operatorname{sig}(x)=1/(1+e^{-x})$ is the logistic function, $t_{\max}$ the saturation angle, $\beta_0$ the inflection point, and $\tau_n=\tau\,(n/64)^{\gamma}$ a steepness whose growth with $n$ is set by $\gamma$. The angle vanishes at $\beta=0$, so the input there is $\ket{+}^{\otimes n}$, and unlike the linear angle $t=\lambda\beta$ of C-IQP (P), it stops growing beyond the training temperatures. The IQP unitary has one-body phases on every qubit and two-body phases on the nearest-neighbor pairs of open rectangular lattices, from $8\times8$ at $64$ qubits to $32\times64$ at $2048$ qubits. At size $n$, the target is $p(x\mid\beta)\propto\exp[\beta(2|x|-n)^2/(2n)]$.

\emph{Training.} We train three fits per size, with the settings of Table~\ref{tab:training_settings}. With the selected budget growth, the masks and reference samples per update grow from $512$ and $1024$ at $64$ qubits to $2880$ and $5824$ at $2048$ qubits. For the parity encoder and the lattice ansatz, an update with $S$ reference samples, $B$ target samples, and $M$ masks of average bodyness $\bar\ell$ requires $O(Mn+Sn+(S+B)M\bar\ell)$ operations, since each amplitude ratio, phase difference, and target character involves only the $O(\bar\ell)$ qubits and generators touched by the mask. The arithmetic cost per update therefore grows no faster than $n^{3/2}$.

\emph{Development score and search.} We select checkpoints on the development grid $\mathcal B_{\mathrm{dev}}=\{0.3,\allowbreak 0.5,\allowbreak \dots,\allowbreak 1.7\}$, which shares no temperature with training or with the MMD test temperatures $\beta\in\{0.4,\allowbreak 0.95,\allowbreak 1.0,\allowbreak 1.05,\allowbreak 1.6\}$. Let $\overline M_n^2(\boldsymbol\theta;\beta)$ denote the squared-MMD estimate at $\beta$, averaged over two independent estimation banks of target, mask, and reference draws and over both bandwidths. We score a checkpoint by
\begin{equation}
E_n(\boldsymbol\theta)=\frac13\left[2\max_{\beta\in\mathcal B_{\mathrm{dev}}}\overline M_n^2(\boldsymbol\theta;\beta)+\frac{1}{1.4}\int_{0.3}^{1.7}\overline M_n^2(\boldsymbol\theta;\beta)\,\mathrm d\beta\right],
\label{eq:scaling-selection-score}
\end{equation}
where the integral uses the trapezoidal rule on $\mathcal B_{\mathrm{dev}}$, and we retain negative unbiased estimates. In a sixteen-qubit study on the $4\times4$ lattice, where exact probabilities are available, this score had a Spearman correlation of $0.97$ with the exact TVD at $40$ test temperatures, aggregated by the same rule, across the $2520$ checkpoints of $120$ random configurations, against $0.83$ for the squared MMD at the training temperatures. The search trains each configuration at $64$ and $256$ qubits and ranks it by the larger of its two scores. We confirm the three leading saturating configurations with three independent fits per size and select the one with the lowest score; Table~\ref{tab:selected_hyperparameters} gives its constants.

\emph{Evaluation.} We draw five MMD test banks independently of training and selection and, every $500$ updates, average the estimated MMD$^{2}$ over the banks, test temperatures, and bandwidths (Fig.~\ref{fig:scaling_observables}(a)). The initial phases are small, so the value at update zero approximates that of the encoder alone. For $\langle m^2\rangle$, we use the selected checkpoint of each fit. At each of $41$ equally spaced temperatures in $[0,2]$, we draw five independent banks of $256$ distinct pairs $\{i,j\}$, estimate each $\langle Z_iZ_j\rangle$ with $4096$ reference draws, and average the resulting unbiased estimates of $\langle m^2\rangle$ over the banks. The target value follows exactly from the one-dimensional magnetization distribution.

\emph{All-to-all connectivity at 64 qubits.} To test the lattice restriction, we ran a separate search of $60$ saturating configurations with the all-to-all two-body ansatz at $64$ qubits, with the estimator budget of the lattice model, and trained three fits of the selected configuration, using the seeds, data, and evaluation of the lattice fits. The all-to-all model removes most of the crossover error and lowers the test MMD$^2$, but underestimates $\langle m^2\rangle$ in the ordered regime beyond the development grid (Table~\ref{tab:scaling_locality64}).

\begin{table}[!htbp]
\centering
\caption{Lattice and all-to-all models at 64 qubits, each with its own selected configuration. Test MMD$^2$ is the held-out value at the selected checkpoint, and $\Delta\langle m^2\rangle$ is the classically estimated model value minus the exact target. Values are means over three fits. Lower absolute values are better.}
\label{tab:scaling_locality64}
\small
\begin{tabular}{lcc}
\toprule
Quantity & $8\times8$ lattice & All-to-all \\
\midrule
Trainable parameters & 176 & 2080 \\
Test MMD$^2$ & 0.00143 & 0.00097 \\
$\Delta\langle m^2\rangle$ at $\beta=0.8$ & $+0.056$ & $-0.027$ \\
$\Delta\langle m^2\rangle$ at $\beta=1.0$ & $+0.097$ & $-0.013$ \\
$\Delta\langle m^2\rangle$ at $\beta=1.2$ & $+0.045$ & $+0.004$ \\
Mean $|\Delta\langle m^2\rangle|$, $0.8\le\beta\le1.3$ & 0.065 & 0.012 \\
$\Delta\langle m^2\rangle$ at $\beta=1.7$ & $-0.010$ & $-0.106$ \\
$\Delta\langle m^2\rangle$ at $\beta=2.0$ & $-0.008$ & $-0.184$ \\
\bottomrule
\end{tabular}
\end{table}

\subsection{Deployment on quantum hardware}
\label{sec:hardware}
\label{sec:appendix_hardware}

We train conditional IQP Born machines at $16$ and $30$ qubits, on $4\times4$ and $5\times6$ lattices, and deploy them on IQM Emerald, a $54$-qubit superconducting processor with square-lattice topology. The $5\times6$ lattice is the largest rectangle of our ansatz that fits on the couplers calibrated on the day of the experiment, and we place each lattice to maximize the product of the calibrated fidelities of the CZ gates and readouts it uses. Each two-body phase compiles to two CZ gates, and the parity encoder to a tree of CNOT gates along lattice couplers, so the circuits need no routing. More generally, the lattice ansatz maps onto the couplers of square-lattice processors \cite{aruteQuantumSupremacyUsing2019,abdurakhimovTechnologyPerformanceBenchmarks2024,ibmNighthawk2025}, whereas earlier hardware deployments of IQP Born machines used heavy-hexagonal connectivity \cite{balloShallowIQP2026,slimIQPCalorimeter2026,ryuDeploymentLargeIQP2026}. The compiled $16$- and $30$-qubit circuits contain $74$ and $152$ CZ gates, with native-gate depths $62$ and $92$. We sample the training temperatures $\beta=0.6$ and $1.4$, the crossover $\beta=1$, and the extrapolated temperature $\beta=2$, with $3000$ shots each. All runs took place on 2026-10-16, and the repository contains the calibration data of that day \cite{balloConditionalGenerationIQP2026}.

\begin{figure*}[t]
    \centering
    \includegraphics[width=\linewidth]{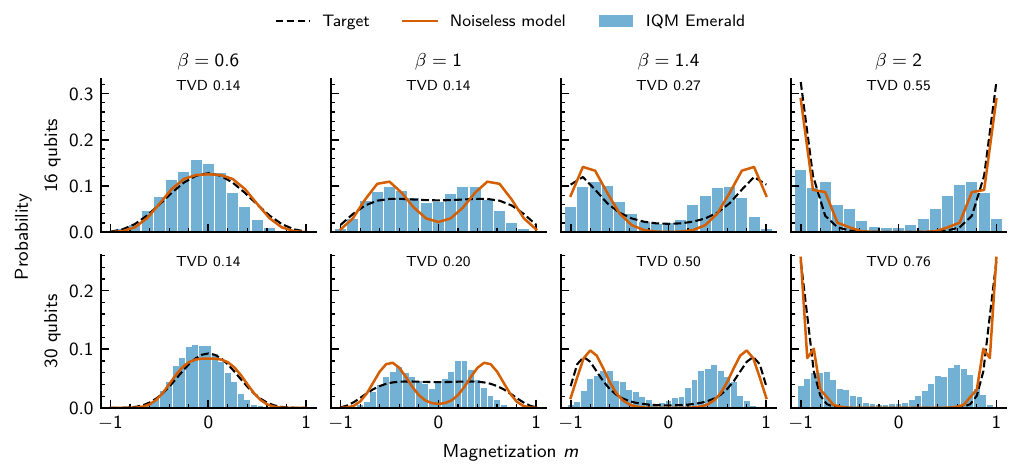}
    \caption{Magnetization profiles sampled from classically trained conditional IQP Born machines on IQM Emerald, without error mitigation. Bars show the distribution of $m=(2|x|-n)/n$ over $3000$ shots at each inverse temperature, dashed lines the exact target, and solid lines the exact profile of the noiseless model. Each panel reports the total variation distance between the hardware and target profiles.}
    \label{fig:hardware_profiles}
\end{figure*}

Figure~\ref{fig:hardware_profiles} shows that the hardware reproduces the transition from a unimodal to a bimodal magnetization profile at both sizes. The exact profiles of the noiseless models, obtained by state-vector simulation, agree with the classical estimates of $\langle m^2\rangle$ in Table~\ref{tab:hardware_m2} to within $0.002$. At $\beta=0.6$, the hardware profiles have TVD $0.14$ to the target at both sizes. At the crossover $\beta=1$, they are closer to the target than the noiseless models, with TVD $0.14$ and $0.20$ against $0.20$ and $0.27$ for $16$ and $30$ qubits, because noise partly fills the dip of the lattice models near $m=0$ (Sec.~\ref{sec:results-scaling}). In the ordered regime, noise moves weight from the ordered peaks toward $m=0$, and the TVD at $\beta=2$ grows to $0.55$ and $0.76$, against $0.14$ for the noiseless models. For $\beta\ge1$, the hardware retains about $60\%$ of the model's excess correlation $\langle m^2\rangle-1/n$ at $16$ qubits and about $47\%$ at $30$ qubits, nearly independently of temperature, a deviation more than an order of magnitude above shot noise. A suppression that is nearly uniform in $\beta$ and stronger for the circuit with about twice as many CZ gates is consistent with gate errors and decoherence. Readout mitigation with M3 \cite{nationScalableMitigationMeasurement2021} changes the profile TVDs by at most $0.02$, so we report unmitigated results.

\begin{table}[t]
\centering
\caption{Mean squared magnetization on IQM Emerald. Hardware values are sample means over $3000$ shots, with one standard error from shot noise. Model values are classical estimates for the noiseless model (Sec.~\ref{sec:scaling}), with one standard error over the five estimation banks of correlator pairs. The last column gives the TVD between the hardware and target magnetization profiles.}
\label{tab:hardware_m2}
\small
\setlength{\tabcolsep}{7pt}
\begin{tabular}{rrcccc}
\toprule
$n$ & $\beta$ & Hardware $\langle m^2\rangle$ & Model $\langle m^2\rangle$ & Target & Profile TVD \\
\midrule
16 & 0.6 & $0.105\pm0.002$ & $0.116\pm0.001$ & 0.131 & 0.139 \\
16 & 1 & $0.217\pm0.004$ & $0.306\pm0.003$ & 0.278 & 0.139 \\
16 & 1.4 & $0.406\pm0.005$ & $0.615\pm0.001$ & 0.574 & 0.269 \\
16 & 2 & $0.533\pm0.005$ & $0.834\pm0.002$ & 0.888 & 0.551 \\
\midrule
30 & 0.6 & $0.058\pm0.001$ & $0.0712\pm0.0003$ & 0.075 & 0.139 \\
30 & 1 & $0.135\pm0.002$ & $0.247\pm0.001$ & 0.205 & 0.195 \\
30 & 1.4 & $0.296\pm0.004$ & $0.588\pm0.001$ & 0.593 & 0.497 \\
30 & 2 & $0.424\pm0.005$ & $0.879\pm0.001$ & 0.904 & 0.765 \\
\bottomrule
\end{tabular}
\end{table}

\clearpage
\end{document}